\documentclass[a4paper,10pt]{article}

\usepackage{bm}
\usepackage{verbatim}
\usepackage[colorlinks=false]{hyperref} 
\usepackage{subfigure}
\usepackage{amsmath,amsfonts,amssymb,graphics,graphicx,epsfig,color,times}
\usepackage{pgfplots}
\usepackage{verbatim}
\usepackage{epstopdf}

\def\picill#1by#2(#3)
{\vbox to #2 {\hrule width #1 height 0pt depth 0pt
\vfill\epsffile{#3}}}

\newcommand{\eq}{\begin{equation}}
\newcommand{\en}{\end{equation}}
\newcommand{\eqa}{\begin{eqnarray}}
\newcommand{\ena}{\end{eqnarray}}

\begin{document}

\setlength{\unitlength}{1mm}

\thispagestyle{empty}


 \begin{center}
{ \bf GHZ transform (I):  Bell transform and quantum teleportation}
  \\[4mm]

{\small Yong Zhang \footnote{yong\_zhang@whu.edu.cn} and Kun Zhang \footnote{kun\_zhang@whu.edu.cn}}\\[3mm]

Center for Theoretical Physics, Wuhan University, Wuhan 430072, P. R. China

School of Physics and Technology, Wuhan University, Wuhan 430072, P. R. China\\[1cm]

\end{center}

\vspace{0.2cm}

\begin{center}
\parbox{13cm}{
\centerline{\small  \bf Abstract}  \noindent

It is well-known that maximally entangled states such as the Greenberger-Horne-Zeilinger (GHZ) states, with the Bell states as the
simplest examples, are widely exploited in quantum information and computation. We study the application of such maximally entangled
states from the viewpoint of the GHZ transform, which is a unitary basis transformation from the product states to the GHZ states.
The algebraic structure of the GHZ transform is made clear and representative examples for it are verified as multi-qubit
Clifford gates. In this paper, we focus on the Bell transform as the simplest example of the GHZ transform and apply it
to the reformulation of quantum circuit model of teleportation and the reformulation of the fault-tolerant construction of
single-qubit gates and two-qubit gates in teleportation-based quantum computation. We clearly show that there exists a natural algebraic
structure called the teleportation operator in terms of the Bell transform to catch essential points of quantum teleportation, and hence
we expect that there would also exist interesting algebraic  structures in terms of the GHZ transform to play important roles in quantum
information and computation.

}

\end{center}

\vspace{.2cm}

\begin{tabbing}

{\bf \small Key Words:} GHZ transform, Bell transform, Teleportation, Bell states,  GHZ states\\[.2cm]

\end{tabbing}


\section{Introduction}

Quantum information and quantum computation \cite{NC2011, Preskill97} is a newly developed research field in which information
processing and computational tasks are accomplished by exploiting fundamental principles of quantum mechanics. Quantum entanglement
\cite{PV05, Bruss02, ZZF00} distinguishes quantum physics from classical physics, and it is widely exploited as a resource in
various topics of quantum information and computation.  The well-known two-qubit maximally entangled states are the Bell states
associated with the Einstein-Podolsky-Rosen paradox \cite{EPR35} or the Bell inequality \cite{Bell64}, and the well-discussed
multi-qubit maximally entangled states are the Greenberger-Horne-Zeilinger (GHZ) states associated with the GHZ theorem \cite{GHZ89, GHZ90}.

An $n$-qubit GHZ transform is defined as a unitary basis transformation from the product basis to the $n$-qubit
GHZ basis which consists of all $2^n$ $n$-qubit GHZ states \cite{GHZ89, GHZ90}. Note that the GHZ
basis allows different forms because these GHZ states can be permuted with each other or can have global phase factors, respectively.
As $n=1$, the GHZ transform is the Hadamard gate \cite{NC2011, Preskill97}. As $n=2$, the GHZ basis is the Bell basis including
four EPR pair states \cite{EPR35,Bell64}, and hence the two-qubit GHZ transform is called  the Bell transform in this paper.

Using quantum entanglements and quantum measurements, quantum teleportation  \cite{BBCCJPW93, Vaidman94, BBC98, BDM00, Werner01,PEWFB15} is an
information protocol of transmitting an unknown qubit from Alice to Bob. Meanwhile, quantum teleportation is a quantum computation
primitive exploited by universal quantum computation called teleportation-based quantum computation \cite{GC99, Nielsen03, Leung04, ZZP15}.
We introduce the Bell transform to characterize the Bell states, and then apply it to the reformulation of the
circuit model of quantum teleportation and further to the reformulation of the fault-tolerant construction of a universal quantum gate set
in teleportation-based quantum computation.

Our proper motivation is to study the nature and the application of quantum maximal entanglement from the viewpoint of
quantum transform, and  it can be stated in two different respects. On the one hand, we characterize the GHZ states with
the GHZ transform so that we can  show how quantum maximal entanglement plays important roles in quantum information and computation in an
algebraic approach. On the other hand, the GHZ transform is regarded as a type of quantum transform in view of the definition and application
of quantum Fourier transform \cite{NC2011, Preskill97}.  Based on the successful application of the Bell transform to quantum teleportation
and teleportation-based quantum computation, we hope that the GHZ transform
would give rise to interesting results in quantum information and computation.

About the GHZ transform, we study its algebraic structure and include known
multi-qubit gates in the literature \cite{FS07,Zhang07} as representative examples. These examples are a higher dimensional
generalization of representative  gates for the Bell transform, and  they are verified as multi-qubit Clifford gates \cite{NC2011, Gottesman97}.
Of course, the GHZ transform may not be a Clifford gate in most cases. We show that the multi-copy of the Pauli $X$ gate can be
obtained as a result of the conjugation by the GHZ transform. Note that a further study of the GHZ transform is
beyond quantum teleportation and teleportation-based quantum computation so it will be submitted elsewhere.

About the Bell transform, we present its state-dependent formulation and matrix formulation,
and collect representative examples for it including the Yang--Baxter gates \cite{YBE67, KL04,ZJG06}, or magic gates proposed by
Makhlin \cite{FS07,Makhlin02}, or matchgates proposed by Valiant \cite{Valiant02, Knill01, JM08, RFSW10,BG11}, or parity-preserving
two-qubit gates \cite{BG11}. These representative gates are recognized as Clifford gates \cite{NC2011, Gottesman97}
and maximally entangling gates \cite{PV05,Bruss02,ZZF00}, but we clearly show that the Bell transform may not be
a Clifford gate in general. Furthermore, we define the teleportation
operator\footnote{The teleportation operator is a direct generalization of the braid teleportation \cite{Zhang06} as
 a tensor product of the identity operator,
the Yang--Baxter gate \cite{YBE67,KL04,ZJG06} and its inverse.} using the Bell transform and  derive the teleportation equation
for the circuit model of quantum teleportation. Moreover,  the fault-tolerant construction \cite{NC2011, Gottesman97} of single-qubit
gates and two-qubit gates in teleportation-based quantum computation can be formulated algebraically using the Bell transform.

Let us claim the findings of our study in the authors' best knowledge. First, we introduce the concept of the GHZ transform to
include various of known two-qubit and multi-qubit quantum gates in the literature. Second, we make clear the algebraic structure of
the GHZ transform  and study its crucial algebraic properties. Third, we introduce the concept of the
teleportation operator and derive the teleportation equation to characterize quantum teleportation and teleportation-based
quantum computation. Fourth, although the quantum circuit of teleportation \cite{BBC98} is usually
viewed as quantum Clifford gate computation \cite{NC2011}, the circuit model of quantum teleportation
using the Bell transform is not because the Bell transform may not be a Clifford gate.

The plan of this  paper is organized as follows. In Section \ref{sec. review Bell basis and quantum gates}, we make a review on
the Bell basis and known quantum gates. In Section \ref{higher_ghz}, we study the algebraic structure of the GHZ
transform with known multi-qubit quantum gates as typical  examples.
In Section \ref{repr_examples}, we define  the Bell transform and study its algebraic structure
with representative examples. In Section \ref{sec: teleportation}, we introduce the teleportation operator using the Bell transform and
then derive the teleportation equation for the characterization of the circuit of quantum teleportation. In Section \ref{fault_tolerant_u_g_s},
we apply the Bell transform to the fault-tolerant construction of
the universal quantum gate set in teleportation-based quantum computation. In Section \ref{sec concluding}, we have concluding remarks.
In Appendix \ref{appendix A}, we show that a permutation gate may not be a Clifford gate. In Appendix \ref{appendix B}, we show that
the representative gates for the Bell transform are maximally entangling Clifford gates.

\section{Review on the  Bell basis and quantum gates}

\label{sec. review Bell basis and quantum gates}

In this section, we set up notations and conventions for the study in the whole paper. We make a short sketch on the product basis and
the Bell basis of the two-qubit Hilbert space. We present a simple review on various of quantum gates in quantum information and
computation \cite{NC2011, Preskill97}, including universal quantum gate sets, Clifford gates, parity-preserving
gates and matchgates, the Yang--Baxter gates, and magic gates.

\subsection{The Bell basis}

A single-qubit Hilbert space  is a two-dimensional Hilbert space ${\mathcal H}_2$, and a two-qubit Hilbert space
is a four-dimensional
Hilbert space ${\cal H}_2\otimes {\cal H}_2$. The orthonormal basis of ${\cal H}_2$ is chosen as the eigenvectors $|0\rangle$ and
$|1\rangle$ of the Pauli matrix $Z$ with $Z|0\rangle=|0\rangle$ and $Z|1\rangle=-|1\rangle$. The Pauli matrices $X$ and $Z$ and the identity
matrix $1\!\! 1_2$ have the conventional form
\eq
 X=\left(\begin{array}{cc}
              0 & 1 \\
              1 & 0 \\
            \end{array} \right), \quad  Z=\left(\begin{array}{cc}
              1 & 0 \\
              0 & -1 \\
            \end{array} \right), \quad 1\!\! 1_2 =\left(\begin{array}{cc}
              1 & 0 \\
              0 & 1 \\
            \end{array}\right).
\en
The product basis of ${\cal H}_2\otimes {\cal H}_2$ denoted by $|k,l\rangle$ or $|kl\rangle$ or $|k\rangle \otimes |l\rangle$ with
$k, l=0,1$, are eigenvectors of the parity-bit operator $Z_1 Z_2$ with $Z_1=Z\otimes 1\!\! 1_2$ and $Z_2=1\!\! 1_2 \otimes Z$:
\eq
  Z_1 Z_2|k,l\rangle \equiv Z\otimes Z |k,l\rangle =(-1)^{k+l}|k,l\rangle,
 \en
where $k+l$ with binary addition modulo 2 represents the parity bit of the state $|k,l\rangle$ and
the lower index of $Z_k$ represents the $k$th qubit Hilbert space. Obviously, $|00\rangle$ and $|11\rangle$ are even-parity states,
while $|01\rangle$ and $|10\rangle$ are odd-parity states.

The Bell states $|\psi(k,l)\rangle$ (or denoted by  $|\psi(kl)\rangle$) \cite{ZZP15} are maximally entangled bipartite pure states 
widely used in quantum information and computation, denoted by
\eq
\label{psi_i-j}
|\psi(k,l)\rangle=(1\!\!1_2\otimes X^l Z^k)|\psi(0,0)\rangle \equiv X_2^l Z_2^k |\psi(0,0)\rangle,
\en
with $|\psi(0,0)\rangle=\frac 1 {\sqrt 2} (|00\rangle + |11\rangle)$ and $k,l=0,1$. For simplicity, the Bell states
$|\psi(k,l)\rangle$ can be described in the other way\footnote{The notation $\widetilde{W}_{\textit{kl}}$ used in this paper is different
from the notation ${W}_{\textit{kl}}$ used in \cite{ZZP15}, and the relationship between them is $\widetilde{W}_{\textit{kl}}=W_{\textit{lk}}$. 
The reason for such the difference is that we want to have a nice formula for the $C_H$ gate  (\ref{ch_gate}) and a nice
formula for the Bell transform (\ref{b_matrix_formulation}).   }
\eq
\label{psi_k-l}
|\psi(k,l)\rangle=(1\!\!1_2\otimes \widetilde{W}_{\textit{kl}})|\Psi\rangle, \quad \widetilde{W}_{\textit{kl}}=X^l Z^k,
\en
with $|\Psi\rangle=|\psi(0,0)\rangle$. They are simultaneous eigenvectors of the parity-bit operator $Z_1 Z_2$ and the
phase-bit operator $X_1 X_2$ given by
 \eqa
 X_1 X_2 |\psi(k,l)\rangle &=& (-1)^k |\psi(k,l)\rangle, \nonumber\\
 Z_1 Z_2 |\psi(k,l)\rangle &=& (-1)^l |\psi(k,l)\rangle,
 \ena
with the phase bit $k$ and the parity bit $l$. The Bell states $|\psi(k,l)\rangle$ give rise to an orthonormal basis of the
two-qubit Hilbert space ${\cal H}_2\otimes {\cal H}_2$, which is called the Bell basis or maximally entangling
basis \cite{NC2011,Preskill97}.

\subsection{Universal quantum gate set}

\label{universal gate set}

Quantum gates  \cite{NC2011, Preskill97} are defined as unitary transformation matrices acting on quantum states, and the set of all $n$-qubit
gates  forms a representation of the unitary group $U(2^n)$.  Both the Hadamard gate $H$ and the \textit{CNOT} gate are often used in the literature
of quantum information and computation \cite{NC2011, Preskill97}, and the Hadamard gate $H$  has the conventional form
\eq
\label{Hadamard gate}
H=\frac 1 {\sqrt 2} (X+Z),
\en
and the \textit{CNOT} gate is defined as
\eq
\label{CNOT gate}
\textit{CNOT}=|0\rangle \langle 0| \otimes 1\!\! 1_2 + |1\rangle\langle 1| \otimes X.
\en
In addition, the \textit{CZ} gate is defined as
\eq
\label{CZ gate}
\textit{CZ} = |0\rangle\langle 0| \otimes 1\!\! 1_2 +  |1\rangle\langle 1| \otimes Z,
\en
and the \textit{CNOT} gate can be related to the  \textit{CZ} gate via the Hadamard gate $H$.

The $T$ gate (the $\pi / 8$ gate \cite{BMPRV00}) has the form
\eq
\label{T_gate}
T=\left(\begin{array}{cc}
1 &  0\\
0 &  e^{i \frac \pi 4}\\
\end{array}\right),
\en
and the set of the Hadamard gate $H$ and the $T$ gate can generate all single-qubit gates.

An entangling two-qubit gate  \cite{ZZF00,BB02} is defined as a two-qubit gate capable of transforming a tensor product of two single-qubit states
into an entangling two-qubit state. For example,  the \textit{CNOT} gate is a maximally entangling gate \cite{BG11}, and it with single-qubit gates
can generate  the Bell states (\ref{psi_i-j}) from the product states.

The set of an entangling two-qubit gate \cite{BB02} with single-qubit gates is called a universal quantum gate set, with which universal quantum
computation  can be performed  in the circuit model  \cite{NC2011} of quantum computation. Hence the set of the \textit{CNOT} gate (or the
\textit{CZ} gate) with single-qubit gates $H$  and $T$  forms a universal quantum gate set.

\subsection{The Clifford gates}

 The set of all tensor products of Pauli matrices \cite{NC2011, Gottesman97} acting on $n$ qubits with phase factors $\pm 1, \pm i$ is called
 the Pauli group ${\mathcal P}_n$. Clifford gates \cite{NC2011, Gottesman97} are defined in two equivalent approaches. They are unitary
 quantum gates preserving tensor products of Pauli matrices under conjugation, or they can be represented as tensor products of the Hadamard
 gate $H$, the phase gate $S$ and the \textit{CNOT} gate. The phase gate $S$ has the form
 \eq
S=\left(\begin{array}{cc}
1 &  0 \\
0 &  i
\end{array}\right),
\en
and obviously $S^2=Z$ and $S^\dag=S^3$ with the Hermitian conjugation $\dag$. The $T$ gate (\ref{T_gate}) is
a square root of the phase gate $S$, namely $S=T^2$, and the transformations of elements of the Pauli
group ${\cal P}_1$ under conjugation by the $T$ gate have the form
\eq
\label{W_clifford}
T X T^\dag =W , \quad T Z T^\dag = Z,
\en
in which $W=\frac {X- i Y} {\sqrt 2}$ is a Clifford gate with the $Y$ gate defined as
\eq
\label{Y_gate}
Y=Z X.
\en
Hence the $T$ gate is not a Clifford gate.

Note that tensor products of the $H$ gate, the $S$ gate and the \textit{CNOT} gate  are only able to
lead to the phase factors $\pm1$ and $\pm i$. Quantum computation of Clifford gates can be efficiently simulated on a classical computer in view of
the Gottesman-Knill theorem \cite{NC2011, Gottesman97}, whereas Clifford gates with the $T$ gate \cite{BMPRV00} are capable
of performing universal quantum computation \cite{NC2011, Preskill97}.

\subsection{The $C_H$ gate}

We define the $C_H$ gate as
\eq
C_H=\textit{CNOT} \cdot H_1,
\en
with $H_1=H \otimes 1\!\! 1_2$, which has the matrix form
\eq
\label{ch_matrix}
C_H=\frac 1 {\sqrt 2}  \left(\begin{array}{cccc}
              1 & 0 & 1 & 0 \\
              0 & 1 & 0 & 1 \\
              0 & 1 & 0 & -1 \\
              1 & 0 & -1 & 0 \\
            \end{array}\right).
\en
Note that the $C_H$ gate is the first example for the Bell transform (or the GHZ transform) in this paper, satisfying
\eq
\label{ch_gate}
|\psi(k,l)\rangle=C_H|k, l\rangle,
\en
which is a unitary transformation from the product basis to the Bell basis.

With the quantum circuits of the $H$ gate and the \textit{CNOT} gate \cite{NC2011, Preskill97},
the associated quantum circuit of the $C_H$ gate is drawn as
\eq
\setlength{\unitlength}{0.6mm}
\begin{array}{c}
\begin{picture}(45,22)
\put(-5,11){$C_H=$}
\put(13,21){\line(1,0){6}}
\put(19,25){\line(1,0){9}}
\put(19,17){\line(0,1){8}}
\put(19,17){\line(1,0){9}}
\put(28,17){\line(0,1){8}}
\put(28,21){\line(1,0){15}}
\put(13,6){\line(1,0){30}}
\put(20.9,19){$H$}
\put(35,3.7){\line(0,1){17.3}}
\put(35,6){\circle{4.8}}
\put(35,21){\circle*{2.1}}
\end{picture}
\end{array},
\en
which is obviously a part of the quantum circuit of quantum teleportation \cite{NC2011,GC99}.

\subsection{Parity-preserving gates and matchgates}

The notation on the parity-preserving gate \cite{Knill01,BG11} refers to our research on quantum computation using the Yang--Baxter gates \cite{KL04,ZJG06},
and it has the form
 \eq \label{matchgate} G(A_G,B_G) =\left(\begin{array}{cccc}
  \omega_1 & 0 & 0 & \omega_7 \\
  0 & \omega_5 & \omega_3 & 0 \\
  0 & \omega_4 & \omega_6 & 0 \\
  \omega_8 & 0 & 0 & \omega_2
  \end{array}\right),\en
with two $SU(2)$ matrices $A_G$ and $B_G$ given by
\eq   A_G =\left(\begin{array}{cc}
  \omega_1 &   \omega_7 \\
  \omega_8 &  \omega_2
  \end{array}\right),\quad  B_G =\left(\begin{array}{cc}
  \omega_5 & \omega_3 \\
  \omega_4 & \omega_6
  \end{array}\right).   \en
The parity-preserving gate $G(A_G,B_G)$  has very good algebraic properties,
\eqa
 G(A_G,B_G)^\dag &=& G(A_G^\dag,B_G^\dag), \nonumber\\
 G(A_G,B_G) G(C_G,D_G) &=& G(A_G C_G,B_G D_G).
\ena
 Note that the $G(A_G,B_G)$ gate is called
the parity-preserving gate because it commutes with the parity-bit operator $Z_1 Z_2$ due to
$Z_1 Z_2 =G(1\!\! 1_2, -1\!\! 1_2)$.

When the determinants of the two $SU(2)$ matrices $A_G, B_G$ are equal, namely $\det(A_G)=\det(B_G)$, the parity-preserving gate $G(A_G,B_G)$  is a matchgate \cite{Knill01,BG11}.
When we call a gate as a parity-preserving gate, we usually mean that it is a parity-preserving non-matchgate. Quantum matchgate computation is associated with
the Valiant theorem \cite{Valiant02,Knill01}, and it can be classically simulated \cite{Valiant02}, and it plays important roles in the research topic \cite{Jozsa08}
of distinguishing classical computation with quantum computation. The set of a matchgate with single-qubit gates \cite{Knill01} is capable of performing universal
quantum computation, and the set of a matchgate with a parity-preserving gate \cite{BG11} can do too.

\subsection{The Yang--Baxter gates $B$ and $B^\prime$}

The Yang--Baxter gates \cite{KL04,ZJG06} are nontrivial unitary solutions of the Yang--Baxter equation \cite{YBE67},
and quantum computation using the Yang--Baxter gates has been explored in recent years.
The Yang--Baxter gate $B$ has the matrix form given by
\eq
\label{b_matrix}
 B=\frac{1}{\sqrt{2}}\left(
                      \begin{array}{cccc}
                        1 & 0 & 0 & 1 \\
                        0 & 1 & -1 & 0 \\
                        0 & 1 & 1 & 0 \\
                        -1 & 0 & 0 & 1 \\
                      \end{array}
                    \right),
 \en
which is the matchgate $B=G(A_B,A_B^{-1})$ with the $SU(2)$ matrix $A_B=e^{\frac \pi 4 Y}$.
The Yang--Baxter gate $B$ is a real orthogonal matrix leading to its inverse and transpose given by
$B^T=B^{-1}=G(A_B^{-1}, A_B)$ which is also a matchgate, with the symbol $T$ denoting the matrix transpose.
Note that the other Yang--Baxter gate $B^\prime$ \cite{Zhang07} given by $B'=G(A_B,A_B)$
has the matrix form
\eq
\label{B'}
 B'=\frac{1}{\sqrt{2}}\left(
                      \begin{array}{cccc}
                        1 & 0 & 0 & 1 \\
                        0 & 1 & 1 & 0 \\
                        0 & -1 & 1 & 0 \\
                        -1 & 0 & 0 & 1 \\
                      \end{array}
                    \right),
 \en
which is a matchgate. Quantum computation of the Yang--Baxter
gate $B$ (or $B^\prime$) can be therefore viewed as an interesting example for quantum matchgate computation \cite{Valiant02,Knill01,JM08,RFSW10,BG11}.

\subsection{Magic gates $Q$ and $R$}

The magic gates are discussed in \cite{Makhlin02, FS07}. With them, tensor products of two single-qubit gates, $\textit{SU(2)}\otimes \textit{SU(2)}$,
can be proved to be isomorphic to the special orthogonal group \textit{SO(4)}. In other words,  two-qubit gates in the special unitary
group \textit{SU(4)} can be characterized by the homogenous space $\textit{SU(4)}/\textit{SO(4)}\otimes \textit{SO(4)}$, namely, two-qubit gates are locally equivalent
when they are associated with single-qubit transformations.

The magic gate $Q$ \cite{Makhlin02}  has the matrix form
 \eq
 \label{q_matrix}
 Q=\frac{1}{\sqrt{2}}\left(
                      \begin{array}{cccc}
                        1 & 0 & 0 & i \\
                        0 & i & 1 & 0 \\
                        0 & i & -1 & 0 \\
                        1 & 0 & 0 & -i \\
                      \end{array}
                    \right),
 \en
and it is the matchgate $Q=G(A_Q, B_Q)$ with two single-qubit gates $A_Q$ and $B_Q$ given by
\eq
A_Q=H\, S, \quad B_Q=i\, A_Q \, Z.
\en
The magic gate $R$  \cite{FS07} has the matrix form
 \eq
 \label{r_matrix}
 R=\frac{1}{\sqrt{2}}\left(
                      \begin{array}{cccc}
                        1 & 0 & 0 & -i \\
                        0 & -i & -1 & 0 \\
                        0 & -i & 1 & 0 \\
                        1 & 0 & 0 & i \\
                      \end{array}
                    \right).
 \en
It is the parity-preserving gate $R=G(A_R, B_R)$ with single-qubit gates
 \eq
 A_R=-i\, B_Q, \quad B_R=-B_Q,
 \en
which give rise to $R=Q\cdot G(Z,-1\!\!1_2)$. Obviously $\det(A_R)\neq \det (B_R)$, so the $R$ gate is a non-matchgate.

\section{The GHZ transform}

\label{higher_ghz}

We define the GHZ transform $\textit{GHZ}^{(n)}$ as a unitary basis transformation
from the $n$-qubit product basis to  the $n$-qubit GHZ basis \cite{GHZ89, GHZ90, FS07, Zhang07}. Representative examples
for it are the higher dimensional generalizations of the $C_H$ gate (\ref{ch_matrix}), the Yang--Baxter gates $B$ (\ref{b_matrix}) 
and $B'$ (\ref{B'}), and the magic gates $Q$ (\ref{q_matrix})
and $R$ (\ref{r_matrix}) in Section~\ref{sec. review Bell basis and quantum gates}, and they are respectively denoted by the $C_H^{(n)}$ 
gate, the Yang--Baxter gates $B^{(n)}$ and  $B'^{(n)}$, and the magic gate $R^{(n)}$. We verify these examples as multi-qubit Clifford
gates \cite{NC2011, Gottesman97}, and with them study the multi-copy of the Pauli $X$ gate.

\subsection{Review on the GHZ basis}

In the stabilizer formalism \cite{NC2011,Gottesman97}, an $n$-qubit GHZ state $|G(j_1,j_2,\ldots,j_n)\rangle$
is specified as an eigenstate of the phase-bit operator $X_1 X_2 \ldots X_n$, the first parity-bit operator $Z_1 Z_2$,
the $i-1$th parity-bit operator $Z_{i-1} Z_i$,  namely
\eqa
X_1X_2\ldots X_n|G(j_1,j_2,\ldots,j_n)\rangle &=& (-1)^{j_1}|G(j_1,j_2,\ldots,j_n)\rangle,\\
Z_1 Z_2|G(j_1,j_2,\ldots,j_n)\rangle &=& (-1)^{j_2}|G(j_1,j_2,\ldots,j_n)\rangle,\\
Z_{i-1}Z_{i}|G(j_1,j_2,\ldots,j_n)\rangle &=& (-1)^{j_i+j_{i-1}}|G(j_1,j_2,\ldots,j_n)\rangle,
\ena
where $j_1$ stands for the phase bit, $j_2$ for the first parity bit, and $j_i+j_{i-1}$ with binary addition for the $i-1$th
parity bit, $3\le i \le n$.

In the $n$-qubit Hilbert space, there are $2^n$ GHZ states $|G(j_1,j_2,\ldots,j_n)\rangle$  which form an orthonormal basis
called the GHZ basis. An $n$-qubit GHZ state in the GHZ basis has the conventional form
\eq
\label{G_J}
|G_J(j_1,j_2,\ldots,j_n)\rangle=\frac{1}{\sqrt{2}}(|0j_2\ldots j_n\rangle+(-1)^{j_1}|1\overline{j}_2\ldots \overline j_n\rangle),
\en
where $j_k=0,1$ and $j_k+\overline j_k=1$ with binary addition, $1\le k \le n$. The subscript $J$ given by
\eq
\label{J}
J(j_1,j_2,\ldots,j_n)=2^{n-1}\cdot j_1+2^{n-2}\cdot j_2+\ldots +2\cdot j_{n-1}+j_n+1,
\en
with decimal addition denotes the GHZ states in a concise way, $1\le J \le 2^n$.

For example, the GHZ basis in the three-qubit Hilbert space has the form
\eqa
\label{G_1}
&&|G_1\rangle = \frac{1}{\sqrt{2}}(|000\rangle+|111\rangle);\quad |G_5\rangle=\frac{1}{\sqrt{2}}(|000\rangle-|111\rangle);\nonumber\\
&&|G_2\rangle = \frac{1}{\sqrt{2}}(|001\rangle+|110\rangle);\quad |G_6\rangle=\frac{1}{\sqrt{2}}(|001\rangle-|110\rangle);\nonumber\\
&&|G_3\rangle = \frac{1}{\sqrt{2}}(|010\rangle+|101\rangle);\quad |G_7\rangle=\frac{1}{\sqrt{2}}(|010\rangle-|101\rangle);\nonumber\\
&&|G_4\rangle = \frac{1}{\sqrt{2}}(|011\rangle+|100\rangle);\quad |G_8\rangle=\frac{1}{\sqrt{2}}(|011\rangle-|100\rangle).
\ena

Besides the notation $|G_J\rangle$ (\ref{G_J}) for an $n$-qubit GHZ state, there is the other notation $|\Phi_K\rangle$ in the
literature \cite{FS07, Zhang07} given by
\eq
\label{Phi_K}
|\Phi_K(j_1,j_1+j_2,\ldots,j_1+j_n)\rangle\equiv|G_J(j_1,j_2,\ldots,j_n)\rangle,
\en
where $j_1+j_i$ is binary addition, $i=2,3,\ldots,n$, and the subscript $K$ is defined by
\eq
\label{K}
K(j_1,j_1+j_2,\ldots,j_1+j_n)=2^{n-1}\cdot j_1+2^{n-2}\cdot (j_1+j_2)+\ldots +(j_1+j_n)+1,
\en
with decimal addition. The relation between two subscripts  $J$ (\ref{J}) and $K$ (\ref{K}) is
\eq
K=\left\{\begin{array}{lll} J,&j_1=0,&1\leqslant J \leqslant 2^{n-1};\\ 2^n+2^{n-1}+1-J,&j_1=1,& 2^{n-1}+1\leqslant J\leqslant 2^n. \end{array}\right.
\en
To show the difference between two kinds of notations $|G_J\rangle$ (\ref{G_J}) and $|\Phi_K\rangle$ (\ref{Phi_K})
for the GHZ basis, we present the GHZ basis in the two-qubit Hilbert space,
  \eq
  (|G_1\rangle,|G_2\rangle,|G_3\rangle,|G_4\rangle)=(|\Phi_1\rangle,|\Phi_2\rangle,|\Phi_4\rangle,|\Phi_3\rangle),
  \en
and the GHZ basis in the three-qubit Hilbert space,
  \eq
  \begin{split}
  &(|G_1\rangle,|G_2\rangle,|G_3\rangle,|G_4\rangle,|G_5\rangle,|G_6\rangle,|G_7\rangle,|G_8\rangle)\\
  =&(|\Phi_1\rangle,|\Phi_2\rangle,|\Phi_3\rangle,|\Phi_4\rangle,|\Phi_8\rangle,|\Phi_7\rangle,|\Phi_6\rangle,|\Phi_5\rangle).
  \end{split}
  \en

\subsection{The definition of the GHZ transform}

\begin{table*}
\begin{center}
\footnotesize
\begin{tabular}{c|c|c}
\hline\hline
Operation & Input & Output \\ \hline
  & $X_1$ & $Z_1$ \\
  & $X_2$ & $X_2$ \\
  & $X_3$ & $X_3$ \\
  & $\vdots$ & $\vdots$ \\
  & $X_n$ & $X_n$ \\  \cline{2-3}
  \raisebox{1.8ex}[0pt]{$C_H^{(n)}$} & $Z_1$ & $X_1X_2\ldots X_n$ \\
  & $Z_2$ & $Z_1Z_2$ \\
  & $Z_3$ & $Z_1Z_3$ \\
  & $\vdots$ & $\vdots$ \\
  & $Z_n$ & $Z_1Z_n$ \\  \hline\hline
\end{tabular}
\end{center}
\caption{\label{C_H Pauli} Transformation properties of elements of the Pauli group ${\mathcal P}_n$ under conjugation by the $C_H^{(n)}$ gate (\ref{C_H_n}).
For example, $C_H^{(n)} X_1 (C_H^{(n)})^\dag=Z_1$. }
\end{table*}

A higher dimensional generalization of the $C_H$ gate (\ref{ch_matrix}), denoted as $C_H^{(n)}$, represents the unitary basis transformation
 matrix from the $n$-qubit product states $|j_1,j_2,\ldots,j_n\rangle$ to the $n$-qubit GHZ states $|G(j_1,j_2,\ldots,j_n)\rangle$ (\ref{G_J}). It is expressed as
\eq
|G(j_1,j_2,\ldots,j_n)\rangle=C_H^{(n)}|j_1,j_2,\ldots,j_n\rangle,
\en
 so the $C_H^{(1)}$ gate is the Hadamard gate $H$ (\ref{Hadamard gate}) and the $C_H^{(2)}$ gate  is the $C_H$ gate (\ref{ch_matrix}). The $C_H^{(n)}$ gate has the form
 as a tensor product of the Hadamard gate $H$ and
the \textit{CNOT} gates,
\eq
\label{C_H_n}
C_H^{(n)}=\textit{CNOT}_{1,n}\textit{CNOT}_{1,n-1}\ldots \textit{CNOT}_{1,2}H_1,
\en
in which the $\textit{CNOT}_{\textit{ij}}$ gate denotes the \textit{CNOT} gate with qubit at site $i$ as the control and qubit at site $j$
as the target. Therefore the $C_H^{(n)}$ gate is a Clifford gate obviously. With the notations (\ref{G_J}) and (\ref{Phi_K}) of the GHZ basis,
the $C_{H}^{(n)}$ gate has the forms given by
\eqa
\begin{split}
C_H^{(n)} &= (|G_1\rangle,|G_2\rangle,\ldots,|G_{2^n}\rangle)  \\
   &= (|\Phi_1\rangle,|\Phi_2\rangle,\ldots,|\Phi_{2^n}\rangle,|\Phi_{2^n-1}\rangle,\ldots,|\Phi_{2^{n-1}+1}\rangle).
   \end{split}
\ena
The transformation properties of elements of the Pauli group ${\mathcal P}_n$ under conjugation by the $C_H^{(n)}$ gate are shown
in Table~\ref{C_H Pauli}.

We define the state-dependent formulation of the GHZ transform as
\eq
\label{GHZ_transform}
\textit{GHZ}^{(n)}=\sum_{j_1,j_2,\ldots,j_n=0}^1\, e^{i\phi_{\textit{k}_1\textit{k}_2\ldots \textit{k}_n}}\, |G(k_1,k_2,\ldots,k_n)\rangle\langle j_1,j_2,\ldots,j_n|,
\en
because there is a bijective mapping between the product states $|j_1,j_2,\ldots,j_n\rangle$ and GHZ states $|G(k_1,k_2,\ldots,k_n)\rangle$
modulo global phases $e^{i\phi_{\textit{k}_1\textit{k}_2\ldots \textit{k}_n}}$. In terms of the $C_H^{(n)}$ gate (\ref{C_H_n}), the $n$-qubit permutation gate $P^{(n)}$
and phase gate $E^{(n)}$ given by
\eqa
\label{P_n}
\begin{split}
P^{(n)}&=\sum_{j_1,j_2,\ldots,j_n=0}^1 |k_1,k_2,\ldots, k_n\rangle\langle j_1,j_2,\ldots,j_n|,\\
\label{E n}
E^{(n)}&=\sum_{j_1,j_2,\ldots,j_n=0}^1 e^{i\phi_{\textit{k}_1 \textit{k}_2 \ldots \textit{k}_n}} |j_1,j_2,\ldots,j_n\rangle\langle j_1,j_2,\ldots,j_n|,
 \end{split}
\ena
the GHZ transform $\textit{GHZ}^{(n)}$ has the other form
\eq
\label{high_Bell}
\textit{GHZ}^{(n)}=C_H^{(n)}P^{(n)}E^{(n)},
\en
which clearly shows the algebraic structure of  the GHZ transform.

The GHZ transform  (\ref{high_Bell}) is not a Clifford gate in general.
The $n$-qubit ($n\ge 3$) permutation gate (\ref{P_n}) may not be a Clifford gate. For example, the Toffoli gate and the Fredkin gate \cite{NC2011}
are three-qubit permutation gates but they are not Clifford gates (Appendix~\ref{appendix A}). The $n$-qubit
phase gate $E^{(n)}$ (\ref{E n}) is not a Clifford gate when the phase factors $e^{i\phi_{\textit{k}_1\textit{k}_2\ldots \textit{k}_n}}$ are not $\pm 1$ or $\pm i$.
Furthermore, the GHZ transform (\ref{high_Bell}) is a maximally entangling multi-qubit gate, in view of the fact that the GHZ states \cite{GHZ89, GHZ90}
are always chosen as maximally entangling multi-qubit states in various entanglement theories \cite{PV05, Bruss02}.

\subsection{The higher dimensional Yang--Baxter gates $B^{(n)}$ and ${B^\prime}^{(n)}$}

The Yang--Baxter gates $B^{(n)}$ and ${B'}^{(n)}$ are the higher dimensional generalization of the four-dimensional Yang--Baxter gates $B$ (\ref{b_matrix}) and $B'$ (\ref{B'}), respectively,
and they satisfy the generalized Yang--Baxter equation \cite{Zhang07}. The $B^{(n)}$ gate is
given by
\eq
\label{B_n}
B^{(n)}=e^{\frac{\pi}{4}M_n},\quad M_n=X^{\otimes n-1}\otimes Y,
\en
with the $Y$ gate defined in (\ref{Y_gate}), and the  $B'^{(n)}$ gate is given by
\eq
\label{B_n'}
{B'}^{(n)}=e^{\frac{\pi}{4}M_n'},\quad M_n'=Y\otimes X^{\otimes (n-1)}.
\en

The $n$-qubit Yang--Baxter gate $B^{(n)}$ is the GHZ transform expressed as
\eq
B^{(n)}=C_H^{(n)}P^{(n)}_{B}E^{(n)}_{B},
\en
with the permutation gate $P^{(n)}_{B}$ and the phase gate $E^{(n)}_B$ given by
\eqa
\begin{split}
P^{(n)}_{B}&=\sum_{j_1,j_2,\ldots,j_n=0}^1|j_n+1,j_1+j_2,\ldots,j_1+j_n\rangle\langle j_1,j_2,\ldots,j_n|,\\
E^{(n)}_B&=\sum_{j_1,j_2,\ldots,j_n=0}^1(-1)^{j_1\cdot (j_n+1)}|j_1,j_2,\ldots,j_n\rangle\langle j_1,j_2,\ldots,j_n|.
\end{split}
\ena
For example, the two-qubit Yang--Baxter gate $B^{(2)}=B$  has the form
\eqa
\begin{split}
B^{(2)}&=(|\Phi_4\rangle,|\Phi_2\rangle,-|\Phi_3\rangle,|\Phi_1\rangle)\\
&=(|G_3\rangle,|G_2\rangle,-|G_4\rangle,|G_1\rangle),
\end{split}
\ena
and the three-qubit Yang--Baxter gate $B^{(3)}$ is given by
\eqa
\begin{split}
B^{(3)}&=(|\Phi_8\rangle,|\Phi_2\rangle,|\Phi_6\rangle,|\Phi_4\rangle,-|\Phi_5\rangle,|\Phi_3\rangle,-|\Phi_7\rangle,|\Phi_1\rangle)\\
&=(|G_5\rangle,|G_2\rangle,|G_7\rangle,|G_4\rangle,-|G_8\rangle,|G_3\rangle,-|G_6\rangle,|G_1\rangle).
\end{split}
\ena
The $B^{(n)}$ gate is an $n$-qubit Clifford gate, and the transformation properties of
elements of the Pauli group ${\mathcal P}_n$ under conjugation by  $B^{(n)}$ are shown  in Table~\ref{B n Pauli}.
\begin{table*}
\begin{center}
\footnotesize
\begin{tabular}{c|c|c}
\hline\hline
Operation & Input & Output \\ \hline
  & $X_1$ & $X_1$ \\
  & $X_2$ & $X_2$ \\
  & $X_3$ & $X_3$ \\
  & $\vdots$ & $\vdots$ \\
  & $X_{n-1}$ & $X_{n-1}$ \\
  & $X_n$ & $X_1X_2\ldots X_{n-1}Z_n$ \\  \cline{2-3}
  \raisebox{1.8ex}[0pt]{$B^{(n)}$} & $Z_1$ & $-Y_1X_2X_3\ldots X_{n-1} Y_n$ \\
  & $Z_2$ & $-X_1Y_2X_3\ldots X_{n-1} Y_n$ \\
  & $Z_3$ & $-X_1X_2Y_3\ldots X_{n-1} Y_n$ \\
  & $\vdots$ & $\vdots$ \\
  & $Z_{n-1}$ & $-X_1X_2X_3\ldots Y_{n-1} Y_n$ \\
  & $Z_n$ & $-X_1X_2X_3\ldots X_{n-1} X_n$ \\  \hline\hline
\end{tabular}
\end{center}
\caption{\label{B n Pauli} Transformation properties of elements of the Pauli group ${\mathcal P}_n$ under conjugation by the $B^{(n)}$ gate (\ref{B_n}).
For example, $B^{(n)} X_1  (B^{(n)})^\dag =X_1$. Note that the $Y$ gate is defined in (\ref{Y_gate}). }
\end{table*}

The higher dimensional Yang--Baxter gate ${B^\prime}^{(n)}$ (\ref{B_n'}) is expressed as
\eq
{B'}^{(n)}=C_H^{(n)}P^{(n)}_{B'}E^{(n)}_{B'},
\en
with the permutation gate $P^{(n)}_{B'}$ and the phase gate $E^{(n)}_{B'}$ given by
\eqa
\begin{split}
P^{(n)}_{B'}&=\sum_{j_1,j_2,\ldots,j_n=0}^1|j_1+1,j_1+j_2,\ldots,j_1+j_n\rangle\langle j_1,j_2,\ldots,j_n|,\\
E^{(n)}_{B'}&=1\!\!1_{2^n\times 2^n}.
\end{split}
\ena
For example, the two-qubit Yang--Baxter gate $B'^{(2)}=B'$  has the form
\eqa
\begin{split}
{B'}^{(2)}&=(|\Phi_4\rangle,|\Phi_3\rangle,|\Phi_2\rangle,|\Phi_1\rangle)\\
&=(|G_3\rangle,|G_4\rangle,|G_2\rangle,|G_1\rangle),
\end{split}
\ena
and the three-qubit Yang--Baxter gate $B'^{(3)}$ is given by
\eqa
\begin{split}
{B'}^{(3)}&=(|\Phi_8\rangle,|\Phi_7\rangle,|\Phi_6\rangle,|\Phi_5\rangle,|\Phi_4\rangle,|\Phi_3\rangle,|\Phi_2\rangle,|\Phi_1\rangle)\\
&=(|G_5\rangle,|G_6\rangle,|G_7\rangle,|G_8\rangle,|G_4\rangle,|G_3\rangle,|G_2\rangle,|G_1\rangle).
\end{split}
\ena
The $B'^{(n)}$ gate is an $n$-qubit Clifford gate, and the transformation properties of
elements of the Pauli group ${\mathcal P}_n$ under conjugation by  $B'^{(n)}$ are shown  in Table~\ref{B n' Pauli}.
\begin{table*}
\begin{center}
\footnotesize
\begin{tabular}{c|c|c}
\hline\hline
Operation & Input & Output \\ \hline
  & $X_1$ & $Z_1X_2X_3\ldots X_{n-1}X_n$ \\
  & $X_2$ & $X_2$ \\
  & $X_3$ & $X_3$ \\
  & $\vdots$ & $\vdots$ \\
  & $X_{n-1}$ & $X_{n-1}$ \\
  & $X_n$ & $X_n$ \\  \cline{2-3}
  \raisebox{1.8ex}[0pt]{${B'}^{(n)}$} & $Z_1$ & $-X_1X_2X_3\ldots X_{n-1} X_n$ \\
  & $Z_2$ & $-Y_1Y_2X_3\ldots X_{n-1} X_n$ \\
  & $Z_3$ & $-Y_1X_2Y_3\ldots X_{n-1} X_n$ \\
  & $\vdots$ & $\vdots$ \\
  & $Z_{n-1}$ & $-Y_1X_2X_3\ldots Y_{n-1} X_n$ \\
  & $Z_n$ & $-Y_1X_2X_3\ldots X_{n-1} Y_n$ \\  \hline\hline
\end{tabular}
\end{center}
\caption{\label{B n' Pauli} Transformation properties of elements of the Pauli group ${\mathcal P}_n$ under conjugation by the ${B'}^{(n)}$ gate (\ref{B_n'}).
For example, ${B'}^{(n)}  X_2  ({B'}^{(n)})^\dag =X_2 $.    }
\end{table*}

\subsection{The higher dimensional magic gates $R^{(n)}$ and  $R'^{(n)}$}

The higher dimensional generalization of the magic gates $Q$ (\ref{q_matrix}) and $R$ (\ref{r_matrix})  have been studied in \cite{FS07},
and it presents a representative example of the GHZ transform,
\eq
R^{(n)}=\sum_{j_1,j_2,\ldots,j_n=0}^1  e^{i\phi_{K}} |\Phi_K(j_1,j_2,\ldots,j_n)\rangle\langle j_1,j_2,\ldots,j_n|,
\en
where $|\Phi_K\rangle$ is defined in (\ref{Phi_K}). It can be expressed as
\eq
\label{R n}
R^{(n)}=C_H^{(n)}P^{(n)}_R E^{(n)}_R,
\en
with the permutation gate $P^{(n)}_{R}$ and the phase gate $E^{(n)}_{R}$ given by
\eqa
P_R^{(n)} &=&\sum_{j_1,j_2,\ldots,j_n=0}^1  |j_1,j_1+j_2,\ldots,j_1+j_n\rangle\langle j_1,j_2,\ldots,j_n|,\\
\label{ERN}
E_R^{(n)} &=& \sum_{j_1,j_2,\ldots,j_n=0}^1 e^{i\phi_{K}} |j_1,j_2,\ldots,j_n\rangle\langle j_1,j_2,\ldots,j_n| .
\ena

Note that the $R^{(n)}$ gate is not a Clifford gate since the entries of the phase gate $E^{(n)}_{R}$ may
not be $\pm 1$ or $\pm i$.  When the phase gate $E^{(n)}_{R}$ is an identity matrix, however, the $R'^{(n)}$ gate defined as
$R'^{(n)}=C_H^{(n)}P^{(n)}_R$ is an $n$-qubit Clifford gate due to $R'^{(n)}=Z_1{B^\prime}^{(n)}$, and the transformation properties of
elements of the Pauli group ${\mathcal P}_n$ under conjugation by $R'^{(n)}$ are shown  in Table~\ref{R n Pauli}. In addition, the transformation
properties of the elements $Z_i$ of the Pauli group ${\mathcal P}_n$ under conjugation by $R^{(n)}$ are the same as those
under conjugation by $R'^{(n)}$, namely,
\eq
R^{(n)} Z_i (R^{(n)})^\dag = R'^{(n)} Z_i (R'^{(n)})^\dag,
\en
with  $i=1, \ldots, n$, because the $Z_i$ gates are commutative with the phase gate $E^{(n)}_R$ (\ref{ERN}).

\begin{table*}
\begin{center}
\footnotesize
\begin{tabular}{c|c|c}
\hline\hline
Operation & Input & Output \\ \hline
  & $X_1$ & $Z_1X_2X_3\ldots X_{n-1}X_n$ \\
  & $X_2$ & $X_2$ \\
  & $X_3$ & $X_3$ \\
  & $\vdots$ & $\vdots$ \\
  & $X_{n-1}$ & $X_{n-1}$ \\
  & $X_n$ & $X_n$ \\  \cline{2-3}
  \raisebox{1.8ex}[0pt]{$R'^{(n)}$} & $Z_1$ & $X_1X_2X_3\ldots X_{n-1} X_n$ \\
  & $Z_2$ & $Y_1Y_2X_3\ldots X_{n-1} X_n$ \\
  & $Z_3$ & $Y_1X_2Y_3\ldots X_{n-1} X_n$ \\
  & $\vdots$ & $\vdots$ \\
  & $Z_{n-1}$ & $Y_1X_2X_3\ldots Y_{n-1} X_n$ \\
  & $Z_n$ & $Y_1X_2X_3\ldots X_{n-1} Y_n$ \\  \hline\hline
\end{tabular}
\end{center}
\caption{\label{R n Pauli} Transformation properties of elements of the Pauli group ${\mathcal P}_n$ under conjugation by $R'^{(n)}=C_H^{(n)} P_R^{(n)}$ (\ref{R n}).
For example, $R'^{(n)}  X_2 (R'^{(n)})^\dag =X_2  $.  }
\end{table*}

\subsection{The multi-copy of the Pauli $X$ gate using the GHZ  transform}

In Table~\ref{C_H Pauli}, there is an interesting result given by
\eq
C_H^{(n)} Z_1 (C_H^{(n)})^\dag = X_1 X_2 \ldots X_n,
\en
so the multi-copy of the Pauli $X$ gate \cite{FS07} can be specified as
\eq
(C_H^{(n)} H_1) X_1 (C_H^{(n)} H_1)^\dag = X_1 X_2 \ldots X_n,
\en
where $H_1 Z_1 H_1 =X_1$ is exploited. Note that $C_H^{(n)} H_1$ is a tensor product of \textit{CNOT} gates.
As the higher dimensional permutation gate $P^{(n)}$ (\ref{P_n}) is given by
\eq
P^{(n)}=\sum_{j_1,j_2,\ldots,j_n=0}^1|j_i+l,k_2,\ldots,k_n\rangle\langle j_1,j_2,\ldots,j_n|,
\en
with $l=0, 1$, the GHZ transform  (\ref{high_Bell}) has the property given by
\eq
\textit{GHZ}^{(n)}Z_i \textit{GHZ}^{(n)\dag}=(-1)^l  X_1X_2\ldots X_n,
\en
so we have to introduce $\textit{GHZ}^{(n)} H_i$ instead of the
GHZ transform itself  to obtain the multi-copy of the Pauli $X$ gate.

For example, when the GHZ transform  is
the Yang--Baxter gate $B^{(n)}$ (\ref{B_n}), we have
 \eq
  B^{(n)}Z_nB^{(n)\dag}=-X_1X_2\ldots X_n,
  \en
 in Table~\ref{B n Pauli};  when the GHZ  transform  is the Yang--Baxter
gate $B'^{(n)}$ (\ref{B_n'}), we have
  \eq
  B'^{(n)} Z_1 B'^{(n)\dag}=-X_1X_2\ldots X_n,
  \en
in Table~\ref{B n' Pauli};  when the GHZ transform  is the magic gate $R^{(n)}$ (\ref{R n}),
 \eq
  R^{(n)}Z_1R^{(n)\dag}=X_1X_2\ldots X_n,
  \en
in Table~\ref{R n Pauli}.
Moreover, when the permutation gate  $P^{(n)}$ (\ref{P_n}) is the Fredkin gate or the Toffoli
gate or their higher dimensional generalizations (Appendix~\ref{appendix A}), the multi-copy of the Pauli $X$ gate can be also done
with the GHZ transform (\ref{high_Bell}) which  may not be a multi-qubit Clifford gate. We hope that the multi-copy operation of the Pauli $X$ gate
under the conjugation by the GHZ transform can play the roles in quantum information and computation, as the authors of the reference \cite{FS07}
had stated before.

\section{The Bell transform is the simplest example for the GHZ transform}

\label{repr_examples}

In this section, we study the algebraic structure of the Bell transform and collect representative examples for it.
These examples are Clifford gates \cite{NC2011, Gottesman97}, yet the Bell transform may not be a Clifford gate
in general. Furthermore, we discuss an intuitive classification of the Bell transform.

\subsection{Definition of the Bell transform}

The Bell transform is defined as a unitary basis transformation matrix from the product basis $|k^\prime,l^\prime\rangle$ to the Bell basis
$ e^{i \phi_{\textit{kl}}} |\psi(k,l)\rangle$ with the global phase factor $e^{i \phi_{\textit{kl}}}$, where $k$ and $l$ are bijective
functions $k(k^\prime,l^\prime)$ and $l(k^\prime,l^\prime)$) of $k^\prime, l^\prime$, respectively, so the Bell transform is a bijective
mapping between $|k^\prime,l^\prime\rangle$ and $ e^{i \phi_{\textit{kl}}} |\psi(k,l)\rangle$ given by
\eq
\label{def_bell}
  e^{i\phi_{\textit{kl}}} |\psi(k,l)\rangle =B_{\textit{ell}} |k^\prime, l^\prime\rangle,
\en
where the notation $B_{\textit{ell}}$ denotes the Bell transform. The state-dependent formulation  (\ref{def_bell}) of the Bell transform
gives rise to its matrix form,
\eq
\label{b_matrix_formulation}
B_{\textit{ell}}=\sum_{k^\prime, l^\prime=0}^1e^{i\phi_{\textit{kl}}}|\psi(k,l)\rangle\langle k^\prime, l^\prime|,
\en
with the help of $C_H$ gate (\ref{ch_matrix}), which can be reformulated as
\eq
B_{\textit{ell}}=\sum_{k^\prime, l^\prime=0}^1e^{i\phi_{\textit{kl}}}C_H|k,l\rangle\langle k^\prime, l^\prime|.
\en
Using the permutation gate $P$ and the phase gate $E$ given by
\eq
\label{phase_permutation}
 P=\sum_{k^\prime, l^\prime=0}^1|k,l\rangle\langle k^\prime, l^\prime|,\quad
 E=\sum_{k^\prime, l^\prime=0}^1e^{i\phi_{\textit{kl}}}|k^\prime, l^\prime\rangle\langle k^\prime, l^\prime|,
\en
the Bell transform has a concrete form given by
\eq
\label{b_transform}
B_{\textit{ell}}=C_H PE.
\en
Hence any  Bell transform can be expressed as a product of the $C_H$ gate, the phase gate $E$
and the permutation gate $P$. For example,  when the $P$ and $E$ gates are identity gates, the Bell transform
is the $C_H$ gate.

\subsection{Representative examples for the Bell transform}

In view of the formalism of the Bell transform (\ref{def_bell}) or (\ref{b_transform}),
it is capable of including various of examples in the literature.  Representative examples for the Bell transform
in this paper include the $C_H$ gate (\ref{ch_matrix}), the Yang--Baxter gate $B$ (\ref{b_matrix}), and the magic
gates $Q$ (\ref{q_matrix}) and $R$ (\ref{r_matrix}).  The $C_H$ gate
is exploited in the definition of the Bell transform. The $B$ gate and the $Q$ gate are matchgates, and
the $R$ gate is a parity-preserving gate. Note that quantum computation of matchgates (or parity-preserving gates)
has been well studied in \cite{Valiant02,Knill01,JM08,RFSW10,BG11}.

The Yang--Baxter gate $B$ (\ref{b_matrix}) is the Bell transform because of
\eq
|\psi(l+1,k+l)\rangle = (-1)^{(k+l)\cdot (l+1)} B |k,l\rangle,
\en
with the multiplication $(k+l)\cdot (l+1)$ as the logical AND operation between $k+l$ and $l+1$, and it
has the form of $B=C_H P_B E_B$ with the permutation gate $P_B$  and the phase gate $E_B$ respectively given by
\eq
P_B=\left(
      \begin{array}{cccc}
        0 &0 & 0 &1 \\
        0 &1 & 0 & 0 \\
        1& 0 & 0 & 0 \\
        0 & 0 & 1 & 0 \\
      \end{array}
    \right), \quad
  E_B=\left(
      \begin{array}{cccc}
        1 & 0 & 0 & 0 \\
        0 & 1 & 0 & 0 \\
        0 & 0 &-1 & 0 \\
       0 & 0 & 0 & 1 \\
      \end{array}
    \right).
  \en
Note that the inverse of the Yang--Baxter gate $B$, denoted by $B^{-1}$  is also the Bell transform.
The other Yang--Baxter gate $B^\prime$ (\ref{B'}) can be expressed as the form of the Bell transform
with the permutation gate $P_{B'}$ and the phase gate $E_{B'}$ given by
\eq
P_{B'} =\sum_{k,l=0}^1 |k+1, k+l \rangle \langle k,l|, \quad E_{B'} =1\!\! 1_4.
\en

The magic gate $Q$ (\ref{q_matrix}) is the Bell transform since
\eq
|\psi(k,k+l)\rangle = (-i)^l Q |k,l\rangle,
\en
with the imaginary unit $i$, and it has the matrix form  of $Q=C_H P_Q E_Q$ with
\eq
  P_Q=\left(
      \begin{array}{cccc}
        1 &  0 & 0 & 0 \\
        0 & 1 & 0 & 0\\
        0 & 0 & 0 & 1 \\
        0 & 0 & 1 & 0 \\
      \end{array}
    \right), \quad
  E_Q=\left(
      \begin{array}{cccc}
        1 & 0 & 0 & 0 \\
       0 & i & 0 & 0\\
       0 & 0 & 1 & 0 \\
        0 & 0 & 0 & i \\
      \end{array}
    \right).
  \en
The magic gate $R$ (\ref{r_matrix}) is the Bell transform satisfying
\eq
|\psi(k,k+l)\rangle = (i)^k (i)^{k+l} R |k,l\rangle,
\en
and it has the matrix form of $R=C_H P_R E_R$ given by
 \eq
  P_R=\left(
      \begin{array}{cccc}
        1 & 0 & 0 & 0 \\
        0 & 1 & 0 & 0 \\
        0 & 0 & 0 & 1\\
        0 & 0 & 1 & 0  \\
      \end{array}
    \right), \quad
  E_R=\left(
      \begin{array}{cccc}
        1 & 0 & 0 & 0\\
        0 & -i & 0 & 0 \\
        0 & 0 & -1 & 0 \\
        0 & 0 & 0 & -i \\
      \end{array}
    \right).
  \en

In Appendix~\ref{appendix B}, a further study is performed on representative examples for the Bell transform,
which include the  $C_H$ gate, the Yang--Baxter gate $B$, and the magic gates
$Q$ and $R$, and their inverses $C^{-1}_H$, $B^{-1}$,  $Q^{-1}$,  $R^{-1}$. First, these two-qubit
gates are  verified as Clifford gates  \cite{NC2011, Gottesman97} in various equivalent approaches. Second, the entangling
powers \cite{PV05,Bruss02,ZZF00} of these gates are calculated to verify them as maximally entangling gates. Third, the exponential
formulations of the $B$,  $Q$,  $R$ gates with associated two-qubit Hamiltonians are derived.

\subsection{The Bell transform may not be a Clifford gate}

\label{Not_a_clifford}

Generally, the Bell transform (\ref{b_transform}) is not a Clifford gate. The $C_H$ gate is obviously a Clifford gate, and the
permutation gate $P$ (\ref{phase_permutation}) is verified as a Clifford gate in Appendix~\ref{appendix A}.  But
the phase gate $E$ (\ref{phase_permutation}) is a Clifford gate only in a very special case.
The phase gate $E$ is a diagonal matrix in the product basis, and has a natural decomposed expression:
\eq
E=e^{ia_1}e^{ia_2Z_1}e^{ia_3Z_2}e^{ia_4Z_1Z_2},
\en
where the parameters $a_i$ are decided by $\phi_{\textit{kl}}$ (\ref{def_bell}).
  As the matrix entries of the phase gate $E$ are not $\pm 1$ or $\pm i$,
the phase gate $E$ is not a Clifford gate.

For example, we construct the  Bell transform $C_{HT}$ given by
\eq
\label{CHT}
C_{HT}=C_{H} (T\otimes 1\!\! 1_2)=\frac{1}{\sqrt{2}}\left(
           \begin{array}{cccc}
             1 & 0 & e^{i\frac{\pi}{4}} & 0\\
             0 & 1 & 0 & e^{i\frac{\pi}{4}} \\
             0 & 1 & 0 & -e^{i\frac{\pi}{4}} \\
             1 & 0 & -e^{i\frac{\pi}{4}} & 0 \\
           \end{array}
         \right),
\en
with the associated state-dependent formulation given by
 \eq
\label{CT}
|\psi(k,l)\rangle= (e^{-i\frac{\pi}{4}})^k C_{HT}|k,l\rangle.
\en
The phase gate $E=T_1$ is not a Clifford gate since the $T$ gate (\ref{T_gate}) is not, and thus the $C_{HT}$ gate is
not a Clifford gate. On the other hand,  the generators of the Pauli group ${\mathcal P}_2$ on two qubits,
$X_1$, $X_2$, $Z_1$, $Z_2$,   are transformed under conjugation by the $C_{HT}$ gate (\ref{CHT}) in the way
\eq
\begin{array}{ll} C_{HT} X_1 C_{HT}^\dag = Z_1 + i Z_1 X_1 X_2, & C_{HT} X_2 C_{HT}^\dag =X_2, \\
 C_{HT} Z_1 C_{HT}^\dag = X_1X_2, & C_{HT} Z_2 C_{HT}^\dag=Z_1Z_2, \end{array}
\en
where $Z_1 + i Z_1 X_1 X_2$ is not an element of the Pauli group ${\mathcal P}_2$, and hence the $C_{HT}$ gate (\ref{CHT}) is not
a Clifford gate (which is verified again).

\subsection{The classification of the Bell transform}

\begin{table*}
\begin{center}
\footnotesize
  \begin{tabular}{c|c}\hline\hline
   Class of the Bell  transform & Example \\ \hline
   Non-Clifford-and-non-parity-preserving gate & $C_{HT}$ \\ \hline
   Clifford-and-non-parity-preserving gate & $C_H$ \\ \hline
   Clifford-and-parity-preserving gate & $R$ \\ \hline
   Clifford-and-matchgate & $B, B^\prime, Q$ \\ \hline
   Matchgate-and-non-Clifford gate & $B_T$ \\ \hline
   Parity-preserving-and-non-Clifford gate & $R_T$ \\ \hline\hline
  \end{tabular}
  \caption{\label{class}  The classification of all examples  for the Bell  transform in Section~\ref{repr_examples}. With such the
  classification, it is obvious that the Bell  transform may not be a Clifford gate and may not be a matchgate.}
\end{center}
\end{table*}

Besides the above examples for the Bell  transform, including the $C_{H}$ gate (\ref{ch_matrix}), the $C_{HT}$ gate
(\ref{CHT}), the Yang--Baxter gates $B$ (\ref{b_matrix}) and $B'$ ({\ref{B'}), the magic gates $Q$ (\ref{q_matrix}) and $R$ (\ref{r_matrix}),
there are many other examples for the Bell  transform which are not parity-preserving gates
or Clifford  gates. For example, we construct another two Bell transforms $B_T$ and $R_T$ given by
\eq
\label{BT}
B_T=\frac{1}{\sqrt{2}}\left(
                        \begin{array}{cccc}
                          1 & 0 & 0 & e^{i\frac{\pi}{4}} \\
                          0 & 1 & -e^{i\frac{\pi}{4}} & 0 \\
                          0 & 1 & e^{i\frac{\pi}{4}} & 0 \\
                          -1 & 0 & 0 & e^{i\frac{\pi}{4}} \\
                        \end{array}\right), \quad
R_T=\frac{1}{\sqrt{2}}\left(
                        \begin{array}{cccc}
                          1 & 0 & 0 & -ie^{i\frac{\pi}{4}} \\
                          0 & -i & -e^{i\frac{\pi}{4}} & 0 \\
                          0 & -i & e^{i\frac{\pi}{4}} & 0 \\
                          1 & 0 & 0 & ie^{i\frac{\pi}{4}} \\
                        \end{array}
                      \right),
\en
where $B_T=B T_1$ is a matchgate, and $R_T =R T_1$ is a parity-preserving gate, and $B_T$ and $R_T$ are not Clifford gates.
Refer to Table~\ref{class} in which there is a simple classification of all examples for the Bell transform in this section.
This classification aims at making two things clear: the Bell transform may not be a Clifford gate and the Bell
transform may not be a matchgate. Hence the application of the Bell transform to quantum information and computation
is beyond the Gottesman-Knill theorem \cite{NC2011, Gottesman97} associated with quantum Clifford gate computation
and the Valiant theorem \cite{Valiant02,Knill01} associated with quantum matchgate computation.

\section{Quantum teleportation using the Bell transform}

\label{sec: teleportation}

This section explores the application of the Bell transform (\ref{b_transform}) to 
quantum teleportation \cite{BBCCJPW93, Vaidman94, BBC98, BDM00, Werner01,PEWFB15}. We define the teleportation operator \cite{Zhang06} in 
terms of the Bell  transform and then exploit it to  derive the teleportation equation \cite{Zhang06} capable of characterizing the 
standard description of quantum teleportation. Furthermore, we study the diagrammatical representation of the Bell transform to exhibit the 
topological diagrammatical feature of quantum teleportation. As a remark, since the Bell transform may not be a Clifford gate,
the quantum circuit model of teleportation using the Bell transform is beyond quantum Clifford gate computation \cite{NC2011, Gottesman97}.

\subsection{Review on quantum teleportation}

Quantum teleportation  is an information protocol with which an unknown
qubit is sent from Alice to Bob by successfully performing the operations including state preparation, Bell measurements,
classical communication and unitary correction.

Alice and Bob share the Bell state $|\Psi\rangle$ (\ref{psi_k-l}) and Alice wants to send an unknown qubit $|\alpha\rangle$ to Bob,
namely, they prepare the quantum state $|\alpha\rangle\otimes |\Psi\rangle$ which is reformulated as
\eq
\label{tele_eq111}
|\alpha\rangle \otimes |\Psi\rangle = \frac 1 2 \sum_{i,j=0}^1 |\psi(i,j)\rangle \otimes \widetilde{W}_{\textit{ij}}  |\alpha\rangle,
\en
called the teleportation equation in \cite{Zhang06}. Then, Alice performs the Bell measurements
denoted by $|\psi(ij)\rangle\langle\psi(ij)|$$\otimes 1\!\! 1_2$ on the prepared state
 $|\alpha\rangle\otimes |\Psi\rangle$,  which gives rise to
 \eq
 (|\psi(ij)\rangle\langle\psi(ij)|\otimes 1\!\! 1_2)(|\alpha\rangle\otimes |\Psi\rangle)
 =\frac 1 2 |\psi(ij)\rangle \otimes \widetilde{W}_{\textit{ij}}  |\alpha\rangle,
 \en
and afterward, Alice informs Bob her measurement results labeled as $(i,j)$.
Finally, Bob applies the unitary correction operator $\widetilde{W}^\dag_{\textit{ij}}$ on his state, expressed as
\eq
 (1\!\! 1_2 \otimes 1\!\! 1_2 \otimes \widetilde{W}^\dag_{\textit{ij}}   ) (|\psi(ij)\rangle \otimes \widetilde{W}_{\textit{ij}}  |\alpha\rangle)
  = |\psi(i j)\rangle \otimes |\alpha\rangle,
\en
to obtain the transmitted qubit $|\alpha\rangle$.

\subsection{Quantum teleportation using the $C_H$ gate}

\begin{figure}
  \begin{center}
  \includegraphics[width=8cm]{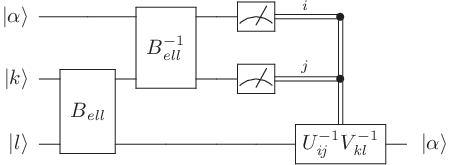}
  \end{center}
  \caption{\label{tele_circuit} Quantum circuit for quantum teleportation using the Bell transform, as a diagrammatical representation of the teleportation equation
  (\ref{tele_eq333}) in terms of the teleportation operator (\ref{tele_operator}). The diagram is read from the left to the right. The single lines denote
  qubits and the double lines denote classical bits. The box $B_{\textit{ell}}$ (or $B^{-1}_{\textit{ell}}$) denotes a two-qubit gate as the Bell transform $B_{\textit{ell}}$
  (or its inverse $B^{-1}_{\textit{ell}}$). Alice has an unknown qubit state $|\alpha\rangle$ and wants to transfer it to Bob, and she makes the Bell measurements on her
  two-qubit state with the measurement outputs  $(i,j)$. When Bob gets the two-bit message $(i,j)$  from Alice, he performs the local unitary correction
  operator $U_{\textit{ij}}^{-1}V_{\textit{kl}}^{-1}$ on his quantum state to obtain $|\alpha\rangle$. Note that the examples for
   $U_{\textit{ij}}$ and $V_{\textit{kl}}$  are shown in Table~\ref{tele_uv}.
 }
  \end{figure}

With the Bell transform $C_H$ (\ref{ch_matrix}), the teleportation
equation (\ref{tele_eq111}) has the form given by
\eq
(1\!\! 1_2 \otimes C_H)  |\alpha\rangle |00\rangle =(C_H \otimes 1\!\! 1_2)\frac 1 2 \sum_{i,j=0}^1 |ij\rangle \widetilde{W}_{\textit{ij}}  |\alpha\rangle,
\en
and it has the other more meaningful form
\eq
\label{tele_eq222}
(C_H^{-1} \otimes 1\!\! 1_2) (1\!\! 1_2 \otimes C_H)  |\alpha\rangle  |00\rangle = \frac 1 2 \sum_{i,j=0}^1 |ij\rangle \widetilde{W}_{\textit{ij}} |\alpha\rangle,
\en
in which $1\!\! 1_2 \otimes C_H$ represents an operation of creating the Bell state $|\Psi\rangle$ and  $C_H^{-1} \otimes 1\!\! 1_2$ is associated
with an operation of performing Bell measurements. After Alice informs Bob the classical two bits $(i,j)$, Bob performs the local unitary correction operator $\widetilde{W}^\dag_{\textit{ij}}$ on his qubit to obtain the transmitted qubit $|\alpha\rangle$.

\subsection{Quantum teleportation using the Bell transform}

\label{subsection tele operator}

\begin{table}
\begin{center}
\footnotesize
\begin{tabular}{c|c|c} \hline\hline
  $B_{\textit{ell}}$ & $U_{\textit{ij}}$ & $V_{\textit{kl}}$  \\ \hline
  $C_H$ & $X^jZ^i$ & $X^lZ^k$  \\
  $Q$ & $(-\sqrt{-1})^{j}X^{i+j}Z^{i}$ & $(\sqrt{-1})^lX^{k+l}Z^{k}$  \\ \hline
  $B$ & $Z^{j+1}X^{i+j}$ & $Z^{l+1}X^{k+l}$  \\
  $R$ & $(\sqrt{-1})^{j}Z^{i}X^{i+j}$ & $(-\sqrt{-1})^{l}Z^{k}X^{k+l}$  \\ \hline\hline
\end{tabular}
\caption{\label{tele_uv} Local unitary operators $U_{\textit{ij}}$ and $V_{\textit{kl}}$ with $i, j=0 ,1$ and $k, l=0, 1$ in the teleportation equation
  (\ref{tele_eq333}) (or in Figure~\ref{tele_circuit}) for the Bell transforms $B_{\textit{ell}}=C_H, B, Q, R$. Here  $i+j$ is the binary
  addition modulo 2. The local unitary operators $U_{\textit{ij}}$ and $V_{\textit{kl}}$ for the Bell transforms $C_H$ and $Q$ have the form of products of the Pauli matrix
  $X$ and the Pauli matrix $Z$, while for  the Bell transforms $B$ and $R$ have of the Pauli matrix $Z$ and the Pauli matrix $X$. The symbol $\sqrt{-1}$ is used to denote the
  imaginary unit because the symbol $i$ has been used as an index. }
\end{center}
\end{table}

Through the teleportation equation (\ref{tele_eq222}), we realize that the operator $(C_H^{-1} \otimes 1\!\! 1_2) (1\!\! 1_2 \otimes C_H)$
plays the key role in the  algebraic formulation of quantum teleportation, so we propose the concept of the teleportation operator given by
\eq
\label{tele_operator}
(B^{-1}_{\textit{ell}} \otimes 1\!\! 1_2)( 1\!\! 1_2 \otimes B_{\textit{ell}}),
\en
or given by
\eq
\label{tele_operator2}
( 1\!\! 1_2 \otimes  B^{-1}_{\textit{ell}}  )( B_{\textit{ell}} \otimes 1\!\! 1_2),
\en
in terms of the Bell transform $B_{\textit{ell}}$ (\ref{b_transform}), its inverse $B^{-1}_{\textit{ell}}$ and the identity operator $1\!\! 1_2$.
In the following, we derive the teleportation equations using the above teleportation operators.

Using the formula (\ref{psi_k-l}), the teleportation equation (\ref{tele_eq111}) has a generalized form
\eq
\label{generalized tell equation}
|\alpha\rangle \otimes |\psi(k,l)\rangle = \frac 1 2 \sum_{i,j=0}^1 |\psi(i,j)\rangle \otimes   \widetilde{W}_{\textit{kl}}  \widetilde{W}_{\textit{ij}} |\alpha\rangle,
\en
which is reformulated with the Bell transform (\ref{def_bell}) as
\eq
(1\!\!1_2\otimes B_{\textit{ell}})|\alpha\rangle\otimes e^{-i\phi_{\textit{kl}}}|k'l'\rangle
 =(B_{\textit{ell}}\otimes 1\!\!1_2)\frac 1 2 \sum_{i,j=0}^1 e^{-i\phi_{\textit{ij}}}|i'j'\rangle\otimes  \widetilde{W}_{\textit{kl}} \widetilde{W}_{\textit{ij}} |\alpha\rangle.
\en
Such the equation has  a further simplified form,
\eq
\label{tele eq indices}
(B^{-1}_{\textit{ell}}\otimes 1\!\!1_2)(1\!\!1_2\otimes B_{\textit{ell}})|\alpha\rangle\otimes |k'l'\rangle
 =\frac 1 2 \sum_{i,j=0}^1 |i'j'\rangle\otimes  \widetilde{V}_{\textit{kl}} \widetilde{U}_{\textit{ij}} |\alpha\rangle,
\en
where two single-qubit gates $\widetilde{V}_{\textit{kl}}$ and $\widetilde{U}_{\textit{ij}}$ have the form
\eq
\label{V_W}
\widetilde{V}_{\textit{kl}}=e^{i\phi_{\textit{kl}}} \widetilde{W}_{\textit{kl}},\quad
\widetilde{U}_{\textit{ij}}=e^{-i\phi_{\textit{ij}}} \widetilde{W}_{\textit{ij}},
\en
and the indices  $i$, $j$, $k$ and $l$ are bijective functions  of $i'$, $j'$, $k'$ and $l'$,
respectively, given by
\eq
\label{index_function}
i=f(i',j'),\quad j=g(i',j'),\quad k=f(k',l'),\quad l=g(k',l').
\en

For notational convenience, we rewrite (\ref{tele eq indices}) as
\eq
\label{tele_u_v}
(B^{-1}_{\textit{ell}}\otimes 1\!\!1_2)(1\!\!1_2\otimes B_{\textit{ell}})|\alpha\rangle\otimes |kl\rangle=\frac 1 2 \sum_{i,j=0}^1 |ij\rangle\otimes
 \widetilde{V}_{\textit{k}'\textit{l}'}  \widetilde{U}_{\textit{i}'\textit{j}'} |\alpha\rangle,
\en
in which the bijective mappings between lower indices are given by
\eq
\label{index_function_p}
i'=f(i,j),\quad j'=g(i,j),\quad k'=f(k,l),\quad l'=g(k,l),
\en
with functions $f$ and $g$ defined in (\ref{index_function}). Furthermore, with the notations $V_{\textit{kl}}$ and $U_{\textit{ij}}$, respectively
 defined by
\eq
\label{new_V_W}
V_{\textit{kl}}=\widetilde{V}_{\textit{k}'\textit{l}'}, \quad  U_{\textit{ij}}=\widetilde{U}_{\textit{i}'\textit{j}'},
\en
we have an appropriate form of the teleportation equation given by
\eq
\label{tele_eq333}
(B^{-1}_{\textit{ell}} \otimes 1\!\! 1_2)( 1\!\! 1_2 \otimes B_{\textit{ell}}) |\alpha\rangle |kl\rangle 
  =\frac 1 2 \sum_{i,j=0}^{1} |ij\rangle V_{\textit{kl}} U_{\textit{ij}} |\alpha\rangle,
\en
with $k,l=0,1$, which is to be exploited in the following study.

We draw Figure~\ref{tele_circuit} as a diagrammatical representation of the teleportation equation (\ref{tele_eq333}) in terms of the teleportation
operator (\ref{tele_operator}). As a matter of fact, it is the quantum circuit model of quantum teleportation, in which Bob performs the local unitary
operation $U_{\textit{ij}}^{-1} V_{\textit{kl}}^{-1}$ on his qubit to obtain the transmitted qubit $|\alpha\rangle$. For example, when the Bell 
transform $B_{\textit{ell}}$ is the $C_H$ gate, the Yang--Baxter gate $B$, the matchgate $Q$ and the parity-preserving gate $R$ 
in Section~\ref{sec. review Bell basis and quantum gates}, the explicit forms 
of the associated single-qubit gates $U_{\textit{ij}}$ and $V_{\textit{kl}}$ are collected in  Table~\ref{tele_uv}.

Note that the phase factors of the single-qubit gates $V_{\textit{kl}}$ and $U_{\textit{ij}}$ (\ref{V_W}) and (\ref{new_V_W}) originally come from the phase
gate $E$ (\ref{phase_permutation}) in the matrix formulation of the Bell transform (\ref{b_transform}). In view of the teleportation equation
(\ref{tele_eq333}), both the unitary correction operators  $(V_{\textit{kl}}U_{\textit{ij}})^\dag$ and $(\widetilde{W}_{\textit{kl}}\widetilde{W}_{\textit{ij}})^\dag$ in
quantum teleportation give rise to the same qubit state $|\alpha\rangle$ modulo a global phase. Hence, the phase
gate $E$ does not play physical roles in view of the performance of quantum teleportation. On the other hand, the phase gate $E$ makes sense in
quantum computation. Usually, the phase gate $E$ is not a Clifford gate, for example, $E=T\otimes 1\!\! 1_2$ with the $T$ gate (\ref{T_gate}), refer to
Subsection~\ref{Not_a_clifford}. Therefore the quantum circuit of teleportation can be regarded as quantum non-Clifford gate computation.

To derive another form of the teleportation equation using the teleportation operator (\ref{tele_operator2}),
we start from the teleportation equation expressed as
\eq
|\Psi\rangle \otimes |\alpha\rangle  = \frac 1 2 \sum_{i,j=0}^1 \widetilde{W}^T_{\textit{ij}}|\alpha\rangle\otimes|\psi(i,j)\rangle,
\en
and then exploit the following property of the Bell states (\ref{psi_k-l}),
\eq
|\psi(k,l)\rangle=(\widetilde{W}^T_{\textit{kl}}\otimes 1\!\!1_2)|\Psi\rangle,
\en
to obtain the teleportation equation
\eq
\label{tele_eq444}
( 1\!\! 1_2 \otimes  B^{-1}_{\textit{ell}}  )( B_{\textit{ell}} \otimes 1\!\! 1_2) |kl\rangle |\alpha\rangle =\frac 1 2 \sum_{i,j=0}^{1}  V^T_{\textit{kl}} U^T_{\textit{ij}} |\alpha\rangle |ij\rangle,
\en
which is to be used in the fault-tolerant
construction of two-qubit gates in teleportation-based quantum computation, refer to Figure~\ref{two_tele} in Subsection \ref{tele_based qc}.

\subsection{The diagrammatical representation of the Bell transform}

In view of the recent research \cite{ZZP15} which shows that the quantum circuit model of teleportation 
admits a nice topological diagrammatical representation, we study the topological diagrammatical representation of the
teleportation operator (\ref{tele_operator}) or (\ref{tele_operator2}), which provides a simpler diagrammatical
proof for deriving the teleportation equation (\ref{tele_eq333}) or (\ref{tele_eq444}). In the following, only specific
diagrammatical rules  \cite{ZZP15} are reviewed just enough for the present usage in this subsection (The complete set of
diagrammatical rules is referred to \cite{ZZP15}).

With the single-qubit gate $V_{\textit{kl}}$ (\ref{new_V_W}), the Bell transform (\ref{b_matrix_formulation}) can be expressed as
\eq
\label{v_bell}
B_{\textit{ell}}=\sum_{k,l=0}^1e^{i\phi_{\textit{k}'\textit{l}'}}|\psi(k',l')\rangle\langle kl|=\sum_{k,l=0}^1 (1\!\!1_2\otimes V_{\textit{kl}})|\Psi\rangle\langle kl|.
\en
In view of the diagrammatical rules \cite{ZZP15}, a single vertical line with the
symbol $\triangle$ denotes a covector product state $\langle 0|$, and the one with the action of the Pauli gate $X$ stands for the state
$\langle 1|$; a solid point on the configuration denotes a single-qubit gate; a cup configuration represents the Bell state $|\Psi\rangle$.
The diagrammatical representation of the Bell transform (\ref{b_matrix_formulation}) is pictured as
\eq
\label{diagram Bell transform}
\setlength{\unitlength}{0.6mm}
\begin{array}{c}
\begin{picture}(80,30)

\put(3,14){$B_{\textit{ell}}=\sum_{k,l=0}^1$}

\put(54,18){\line(0,1){12}}
\put(44,18){\line(0,1){12}}
\put(44,18){\line(1,0){10}}
\put(54,24){\circle*{2.}}
\put(56,23){\tiny{$V_{\textit{kl}}$}}

\put(44,0){\line(0,1){10}}
\put(54,0){\line(0,1){10}}
\put(54,6){\circle*{2.}}
\put(55,5){\tiny{$X^l$}}
\put(44,6){\circle*{2.}}
\put(45,5){\tiny{$X^k$}}

\put(42.3,10){\tiny{$\triangle$}}
\put(52.3,10){\tiny{$\triangle$}}

\end{picture}
\end{array},
\en
in which the diagrammatical representation is read from the bottom to the top and its associated algebraical expression (\ref{v_bell})
is read from the right to the left.

With the single-qubit gate $U_{\textit{ij}}$ (\ref{new_V_W}), the inverse of the Bell  transform (\ref{b_matrix_formulation}) has the form
\eq
B^{-1}_{\textit{ell}}=\sum_{i,j=0}^1|ij\rangle\langle\psi(i',j')|e^{-i\phi_{\textit{i}'\textit{j}'}}=\sum_{i,j=0}^1|ij\rangle\langle\Psi|(1\!\!1_2\otimes U^T_{\textit{ij}}).
\en
In accordance with the diagrammatical rules \cite{ZZP15}, a cap configuration denotes
the complex conjugation of the Bell state $|\Psi\rangle$ and a vertical line with the symbol $\nabla$ denotes the state $|0\rangle$.
The inverse of the Bell transform (\ref{b_matrix_formulation}) has the following diagrammatical representation
\eq
\label{diagram inverse Bell transform}
\setlength{\unitlength}{0.6mm}
\begin{array}{c}
\begin{picture}(80,30)

\put(3,14){$B^{-1}_{\textit{ell}}=\sum_{i,j=0}^1$}

\put(54,20){\line(0,1){10}}
\put(44,20){\line(0,1){10}}
\put(54,24){\circle*{2.}}
\put(55,23){\tiny{$X^j$}}
\put(44,24){\circle*{2.}}
\put(45,23){\tiny{$X^i$}}

\put(42.3,18){\tiny{$\nabla$}}
\put(52.3,18){\tiny{$\nabla$}}

\put(44,0){\line(0,1){12}}
\put(54,0){\line(0,1){12}}
\put(44,12){\line(1,0){10}}
\put(54,6){\circle*{2.}}
\put(55.9,5){\tiny{$U^T_{\textit{ij}}$}}

\end{picture}
\end{array},
\en
which is read from the bottom to the top.

With the help of the diagrammatical representations  (\ref{diagram Bell transform}) and  (\ref{diagram inverse Bell transform}),
the teleportation operator (\ref{tele_operator}) has the diagrammatical representation
\eq
  \setlength{\unitlength}{0.5mm}
  \begin{array}{c}
  \begin{picture}(30,65)

  \put(-110,29){$(B^{-1}_{\textit{ell}}\otimes 1\!\!1_2)(1\!\!1_2\otimes B_{\textit{ell}})=\sum_{i,j,k,l=0}^1$}

  \put(14.5,48){\line(0,1){12}}
  \put(2,48){\line(0,1){12}}
  \put(14.5,54){\circle*{2.}}
  \put(15.5,53){\tiny{$X^j$}}
  \put(2,54){\circle*{2.}}
  \put(3,53){\tiny{$X^i$}}

  \put(-.2,45){{\scriptsize$\nabla$}}
  \put(12.3,45){\scriptsize{$\nabla$}}

  \put(14.5,30){\circle*{2.}}
  \put(16,29){\tiny{$U^T_{\textit{ij}}$}}

  \put(14.5,0){\line(0,1){12}}
  \put(27,0){\line(0,1){12}}
  \put(27,6){\circle*{2.}}
  \put(28,5){\tiny{$X^l$}}
  \put(14.5,6){\circle*{2.}}
  \put(15.5,5){\tiny{$X^k$}}

  \put(12.3,12){\tiny{$\triangle$}}
  \put(24.8,12){\tiny{$\triangle$}}

  \put(2,0){\line(0,1){37.5}}
  \put(2,37.5){\line(1,0){12.5}}
  \put(14.5,22.5){\line(0,1){15}}
  \put(27,22.5){\line(0,1){37.5}}
  \put(14.5,22.5){\line(1,0){12.5}}

  \put(27,30){\circle*{2.}}
  \put(28.5,29){\tiny{$V_{\textit{kl}}$}}

  \put(42,29){$=\sum_{i,j,k,l=0}^1$}

  \put(94.5,48){\line(0,1){12}}
  \put(82,48){\line(0,1){12}}
  \put(94.5,54){\circle*{2.}}
  \put(95.5,53){\tiny{$X^j$}}
  \put(82,54){\circle*{2.}}
  \put(83,53){\tiny{$X^i$}}

  \put(79.8,45){{\scriptsize$\nabla$}}
  \put(92.3,45){\scriptsize{$\nabla$}}


  \put(94.5,0){\line(0,1){12}}
  \put(107,0){\line(0,1){12}}
  \put(107,6){\circle*{2.}}
  \put(108,5){\tiny{$X^l$}}
  \put(94.5,6){\circle*{2.}}
  \put(95.5,5){\tiny{$X^k$}}

  \put(92.3,12){\tiny{$\triangle$}}
  \put(104.8,12){\tiny{$\triangle$}}

  \put(82,0){\line(0,1){37.5}}
  \put(82,37.5){\line(1,0){12.5}}
  \put(94.5,22.5){\line(0,1){15}}
  \put(107,22.5){\line(0,1){37.5}}
  \put(94.5,22.5){\line(1,0){12.5}}

  \put(107,54){\circle*{2.}}
 \put(108.5,53){\tiny{$V_{\textit{kl}}U_{\textit{ij}}$}}

  \end{picture}
  \end{array}
  \en
in which the diagrammatical rules \cite{ZZP15} are exploited: the vertical line represents the identity operator $1\!\!1_2$ and
the single-qubit gate $U^T_{\textit{ij}}$ flows  from one branch to the adjacent branch with the transpose operation.  To derive
the teleportation equation (\ref{tele_eq333}) in a diagrammatical approach, we apply the teleportation operator (\ref{tele_operator})
on the prepared state $|\alpha\rangle|kl\rangle$ and then straighten the connected line of the top cap  with the bottom
cup as a sort of topological deformation, so that the unknown qubit $|\alpha\rangle$ with the action of the local unitary
operation $V_{\textit{kl}}U_{\textit{ij}}$  is transmitted.

To complete the study of the diagrammatical representation of the teleportation operator, we draw the associated configuration of the
teleportation operator (\ref{tele_operator2})  as
\eq
  \setlength{\unitlength}{0.5mm}
  \begin{array}{c}
  \begin{picture}(30,65)

  \put(-110,29){$(1\!\!1_2\otimes B^{-1}_{\textit{ell}})(B_{\textit{ell}}\otimes 1\!\!1_2)=\sum_{i,j,k,l=0}^1$}

  \put(14.5,48){\line(0,1){12}}
  \put(27,48){\line(0,1){12}}
  \put(14.5,54){\circle*{2.}}
  \put(15.5,53){\tiny{$X^i$}}
  \put(27,54){\circle*{2.}}
  \put(29.5,53){\tiny{$X^j$}}

  \put(24.8,45){{\scriptsize$\nabla$}}
  \put(12.3,45){\scriptsize{$\nabla$}}

  \put(14.5,30){\circle*{2.}}
  \put(16,29){\tiny{$V_{\textit{kl}}$}}

  \put(14.5,0){\line(0,1){12}}
  \put(2,0){\line(0,1){12}}
  \put(2,6){\circle*{2.}}
  \put(15.5,5){\tiny{$X^l$}}
  \put(14.5,6){\circle*{2.}}
  \put(3,5){\tiny{$X^k$}}

  \put(12.3,12){\tiny{$\triangle$}}
  \put(-.2,12){\tiny{$\triangle$}}

  \put(27,0){\line(0,1){37.5}}
  \put(2,22.5){\line(1,0){12.5}}
  \put(14.5,22.5){\line(0,1){15}}
  \put(2,22.5){\line(0,1){37.5}}
  \put(14.5,37.5){\line(1,0){12.5}}

  \put(27,30){\circle*{2.}}
  \put(28.5,29){\tiny{$U^T_{\textit{ij}}$}}

  \put(41,29){$=\sum_{i,j,k,l=0}^1$}

  \put(94.5,48){\line(0,1){12}}
  \put(107,48){\line(0,1){12}}
  \put(94.5,54){\circle*{2.}}
  \put(95.5,53){\tiny{$X^i$}}
  \put(107,54){\circle*{2.}}
  \put(108,53){\tiny{$X^j$}}

  \put(104.8,45){{\scriptsize$\nabla$}}
  \put(92.3,45){\scriptsize{$\nabla$}}


  \put(94.5,0){\line(0,1){12}}
  \put(82,0){\line(0,1){12}}
  \put(82,6){\circle*{2.}}
  \put(83,5){\tiny{$X^k$}}
  \put(94.5,6){\circle*{2.}}
  \put(95.5,5){\tiny{$X^l$}}

  \put(92.3,12){\tiny{$\triangle$}}
  \put(79.8,12){\tiny{$\triangle$}}

  \put(107,0){\line(0,1){37.5}}
  \put(82,22.5){\line(1,0){12.5}}
  \put(94.5,22.5){\line(0,1){15}}
  \put(82,22.5){\line(0,1){37.5}}
  \put(94.5,37.5){\line(1,0){12.5}}

  \put(82,54){\circle*{2.}}
  \put(64,53){\tiny{$V^T_{\textit{kl}}U^T_{\textit{ij}}$}}

  \end{picture}
  \end{array}
  \en
which naturally gives rise to the teleportation equation (\ref{tele_eq444}).

\section{Teleportation-based quantum computation using the Bell transform}

\label{fault_tolerant_u_g_s}

Teleportation-based quantum computation has been well studied in both algebraic and topological approach in  \cite{GC99, Nielsen03, Leung04, ZZP15}.
Here we present a brief review on the fault-tolerant construction of single-qubit gates and two-qubit gates using quantum teleportation, and then
make a study on the fault-tolerant construction of the universal quantum gate set in teleportation-based quantum computation using the Bell transform.

 \subsection{Review on teleportation-based quantum computation}

\label{clifford_gate_fault_tolerant}

In quantum information and computation \cite{NC2011}, quantum gates $U$ are classified by
\eq
\label{C_k}
C_k \equiv \{ U | U C_{k-2} U^\dag \subseteq C_{k-1} \},
\en
where $C_1$ denotes the Pauli gates and  $C_2$ denotes the Clifford gates. In fault-tolerant quantum computation \cite{NC2011,Preskill97,Gottesman97},
the fault-tolerant construction of Clifford gates including the Pauli gates can be performed in a systematical approach, and the fault-tolerant
construction of non-Clifford gates such as the $T$ gate (\ref{T_gate}) becomes a problem of how to introduce a set of Clifford gates to play the
role of these non-Clifford gates. Teleportation-based quantum computation \cite{GC99} is fault-tolerant quantum computation because it  fault-tolerantly 
prepares a quantum state with the action of a $C_3$ gate and then  fault-tolerantly applies $C_1$ or $C_2$ gates to such the quantum state using the 
teleportation protocol so that this $C_3$ gate can be fault-tolerantly performed.

To fault-tolerantly perform a single-qubit gate $U\in C_k$ (\ref{C_k}) on the unknown qubit state $|\alpha\rangle$, Alice prepares the two-qubit state
$|\Psi_U\rangle$ given by
\eq
\label{Psi_U}
 |\Psi_U\rangle=(1\!\! 1_2\otimes U)|\Psi\rangle,
 \en
and expresses  $|\alpha\rangle\otimes |\Psi_U\rangle$ as
\eq
\label{tele_single0}
|\alpha\rangle\otimes|\Psi_U\rangle=\frac{1}{2}\sum_{i,j=0}^1|\psi(ij)\rangle \otimes R_{\textit{ij}}U|\alpha\rangle,
\en
where the single-qubit gate $R_{\textit{ij}}$ has the form $R_{\textit{ij}}=U\widetilde{W}_{\textit{ij}}U^\dag\in C_{k-1}$ (\ref{C_k}).
Then Alice makes Bell measurements $|\psi(ij)\rangle\langle\psi(ij)|\otimes 1\!\! 1_2$ and informs Bob her measurement results
labeled  by $(i,j)$. Finally, Bob performs the unitary correction operator $R_{\textit{ij}}^\dag \in C_{k-1}$ to attain $U |\alpha\rangle$.
It is obvious that the difficulty of fault-tolerantly performing the single-qubit gate $U\in C_k$ becomes how to
fault-tolerantly prepare the state $|\Psi_U\rangle$ and perform the single-qubit gate $R_{\textit{ij}}^\dag \in C_{k-1}$.

To fault-tolerantly  perform a two-qubit gate \textit{CU}  on two unknown
single-qubit states $|\alpha\rangle$ and $|\beta\rangle$, we prepare a four-qubit entangled state $|\Psi_{\textit{CU}}\rangle$ given by
\eq
\label{Psi_CU}
|\Psi_{\textit{CU}}\rangle=(1\!\! 1_2\otimes \textit{CU} \otimes 1\!\! 1_2)(|\Psi\rangle \otimes |\Psi\rangle),
\en
with the action of the \textit{CU} gate, and reformulate the prepared state $|\alpha\rangle\otimes|\Psi_{CU}\rangle\otimes|\beta\rangle$ as
\eq
\label{tele_two1}
\begin{split}
& |\alpha\rangle\otimes|\Psi_{\textit{CU}}\rangle\otimes|\beta\rangle\\
=&\frac{1}{4}\sum_{i_1,j_1=0}^1\sum_{i_2,j_2=0}^1(1\!\!1_4\otimes
  Q\otimes P\otimes 1\!\!1_4) (|\psi(i_1j_1)\rangle\otimes \textit{CU}|\alpha\beta\rangle\otimes |\psi(i_2j_2)\rangle),
\end{split}
\en
with $1\!\! 1_4=1\!\! 1_2\otimes 1\!\! 1_2$. The single-qubit gates $Q$ and $P$ in the teleportation equation (\ref{tele_two1}) are
calculated by
\eq
\label{Q P}
Q\otimes P=\textit{CU}(\widetilde{W}_{\textit{i}_1\textit{j}_1}\otimes \widetilde{W}^T_{\textit{i}_2\textit{j}_2})\textit{CU}^\dag,
\en
which informs that the $Q$ and $P$  gates (\ref{Q P}) are single-qubit Pauli gates when the \textit{CU} gate is  a Clifford
gate \cite{NC2011, Gottesman97}.  Next, we perform the Bell measurements given by
\eq
\label{six_bell_measurement}
|\psi(i_1 j_1)\rangle\langle\psi(i_1 j_1)|\otimes 1\!\! 1_2\otimes 1\!\! 1_2 \otimes  |\psi(i_2 j_2)\rangle\langle\psi(i_2 j_2)|,
\en
and with the measurement results labeled by $(i_1,j_1)$ and $(i_2,j_2)$, we perform the unitary correction operator, $Q^\dag\otimes P^\dag$, to
obtain the exact action of the \textit{CU} gate on the two-qubit state $|\alpha\rangle \otimes |\beta\rangle$, namely $\textit{CU}|\alpha\beta\rangle$.
Note that the two-qubit  gate \textit{CU} we study here may not be a controlled-operation two-qubit
  gate such as the \textit{CNOT} gate.

\begin{figure}
  \begin{center}
  \includegraphics[width=8cm]{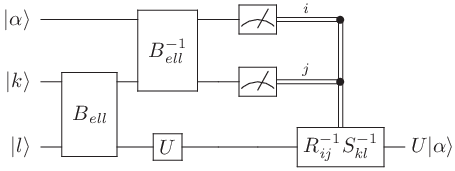}
  \end{center}
 \caption{\label{single_tele} Fault-tolerant construction of the single-qubit gate $U$ in teleportation-based quantum computation using the Bell transform,
 as a diagrammatical representation of the teleportation equation  (\ref{tele_single}) in terms of the teleportation operator (\ref{tele_operator}).
 Alice prepares an unknown qubit state $|\alpha\rangle$ and shares with Bob  a two-qubit state with the local action of the $U$ gate, and Alice makes
 the Bell measurements on her two-qubit system with the measurement outputs $(i,j)$.  When Bob gets the two-bit message $(i,j)$ from Alice,
 he performs the local unitary correction operator $R^{-1}_{\textit{ij}} S_{\textit{kl}}^{-1}$ on his qubit to obtain the qubit state $U|\alpha\rangle$. 
 Note that the examples for the $R_{\textit{ij}}$ and $S_{\textit{kl}}$ gates are shown in Table~\ref{tele_sr_h}
 and Table~\ref{tele_sr_t}.}
  \end{figure}

\begin{table}
\begin{center}
\footnotesize
\begin{tabular}{c|c|c|c}\hline\hline
  $B_{\textit{ell}}$ & $U$ & $R_{\textit{ij}}$ & $S_{\textit{kl}}$  \\ \hline
  $C_H$ & & $Z^jX^i$ & $Z^lX^k$  \\ \cline{1-1} \cline{3-4}
  $B$ & & $X^{j+1}Z^{i+j}$ & $X^{l+1}Z^{k+l}$   \\ \cline{1-1} \cline{3-4}
  $Q$ & \raisebox{1.6ex}[0pt]{$H$} & $(-\sqrt{-1})^{j}Z^{i+j}X^{i}$ & $(\sqrt{-1})^lZ^{k+l}X^{k}$  \\ \cline{1-1} \cline{3-4}
  $R$ & & $(\sqrt{-1})^{j}X^{i}Z^{i+j}$ & $(-\sqrt{-1})^{l}X^{k}Z^{k+l}$  \\ \hline\hline
\end{tabular}
\caption{\label{tele_sr_h} Local unitary operators $R_{\textit{ij}}$ and $S_{\textit{kl}}$ with $i, j=0 ,1$ and $k, l=0, 1$ in the teleportation equation
  (\ref{tele_single}) (or in Figure~\ref{single_tele}) for the Bell transforms $B_{\textit{ell}}=C_H, B, Q, R$ with the single-qubit gate $U$ as the Hadamard gate $H$.
  }
\end{center}
\end{table}

\begin{table}
\begin{center}
\footnotesize
\begin{tabular}{c|c|c|c}\hline\hline
  $B_{\textit{ell}}$ & $U$ & $R_{\textit{ij}}$ & $S_{\textit{kl}}$  \\ \hline
  $C_H$ & & $W^jZ^i$ & $W^lZ^k$  \\ \cline{1-1} \cline{3-4}
  $B$ & & $Z^{j+1}W^{i+j}$ & $Z^{l+1}W^{k+l}$   \\ \cline{1-1} \cline{3-4}
  $Q$ & \raisebox{1.6ex}[0pt]{$T$} & $(-\sqrt{-1})^{j}W^{i+j}Z^{i}$ & $(\sqrt{-1})^l W^{k+l}Z^{k}$  \\ \cline{1-1} \cline{3-4}
  $R$ & & $(\sqrt{-1})^{j}Z^{i}W^{i+j}$ & $(-\sqrt{-1})^{l}Z^{k}W^{k+l}$ \\ \hline\hline
\end{tabular}
\caption{\label{tele_sr_t}  Local unitary operators $R_{\textit{ij}}$ and $S_{\textit{kl}}$ with $i, j=0 ,1$ and $k, l=0, 1$ in the teleportation equation
  (\ref{tele_single}) (or in Figure~\ref{single_tele})  for the Bell  transforms $B_{\textit{ell}}=C_H, B, Q, R$ with the single-qubit gate $U$ as the $T$ gate.
  Note that the single-qubit gate $W$  is the Clifford gate (\ref{W_clifford}).}
\end{center}
\end{table}

\subsection{Teleportation-based quantum computation using the Bell transform}

\label{tele_based qc}

\begin{figure}
  \begin{center}
  \includegraphics[width=8cm]{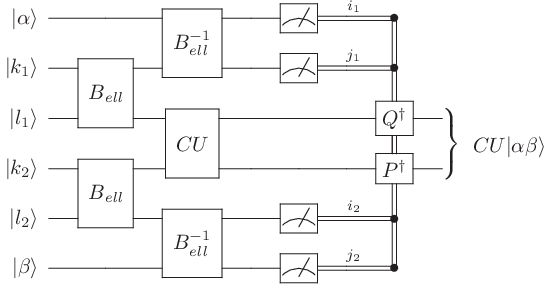}
  \end{center}
  \caption{\label{two_tele} Fault-tolerant construction of the two-qubit gate \textit{CU} in  teleportation-based quantum computation using the
  Bell transform, as a diagrammatical representation of the teleportation equation  (\ref{tele_two}). We prepare an unknown two-qubit
 state $|\alpha\beta\rangle$ and a four-qubit state with the action of the two-qubit gate \textit{CU}, and perform the joint Bell measurements
 with the measurement outputs $(i_1,j_1)$ and $(i_2,j_2)$. With these two-bit messages, we perform the unitary correction operator
 $Q^\dag \otimes P^\dag$ to obtain $|\alpha\beta\rangle$ with the action of the \textit{CU} gate.  Note that the examples for the $Q$ and $P$ gates
 are shown in Table~\ref{tele_qp_cq} and Table~\ref{tele_qp_br}.   }
  \end{figure}

\begin{table*}
\begin{center}
\footnotesize
\begin{tabular}{c|c|c|c}\hline\hline
  $B_{\textit{ell}}$ & \textit{CU} & $Q$ & $P$ \\ \hline
  & \textit{CNOT} & $E_QX^aZ^bZ^c$ & $E_PX^aZ^cX^d$ \\ \cline{2-4}
  & \textit{CZ} & $E_QX^aZ^bZ^d$ & $E_PZ^aZ^cX^d$ \\ \cline{2-4}
  & $C_H$ & $E_QZ^aX^bZ^c$ & $E_PX^bZ^cX^d$\\ \cline{2-4}
  & $C_H^{-1}$ & $E_QZ^aX^bX^c$ & $E_PX^aZ^cX^d$\\ \cline{2-4}
  & $B$ & $(-1)^{b}E_QX^aY^bX^cX^d$ & $(-1)^{c}E_PY^bX^cZ^d$ \\ \cline{2-4}
  \raisebox{1.6ex}[0pt]{$C_H/Q$} & $B^{-1}$ & $E_QX^aY^bX^cX^d$ & $(-1)^dE_PY^bX^cZ^d$ \\ \cline{2-4}
  & $Q$ & $E_QZ^aX^bY^cY^d$ & $(-\sqrt{-1})^dE_PX^aX^bY^cZ^d$ \\ \cline{2-4}
  & $Q^{-1}$ & $(\sqrt{-1})^{a}(\sqrt{-1})^{b}E_QZ^aX^bY^c$ & $(\sqrt{-1})^{c}(\sqrt{-1})^{d}E_PY^aY^bX^cY^d$ \\ \cline{2-4}
  & $R$ & $E_QX^aX^bY^cZ^d$ & $(\sqrt{-1})^dE_PZ^aX^bY^cY^d$ \\ \cline{2-4}
  & $R^{-1}$ & $(-\sqrt{-1})^{a}(-\sqrt{-1})^{b}E_QY^bX^cZ^d$ & $(-\sqrt{-1})^{c}(-\sqrt{-1})^{d}E_PY^aX^bY^cY^d$ \\ \hline\hline
\end{tabular}
\caption{\label{tele_qp_cq}  Local unitary operators $Q$ and $P$  in the teleportation equation
  (\ref{tele_two}) (or in Figure~\ref{two_tele}), with the teleportation operator (\ref{tele_operator}) in terms of the Bell transforms $C_H$ and $Q$,
  for two-qubit gates \textit{CU} including the \textit{CNOT} gate, the \textit{CZ} gate, the Bell transforms $B_{\textit{ell}}=C_H, B, Q, R$ and their inverses.
 The local unitary gates $Q$ and $P$ have the forms as products of the Pauli matrices
  $X$ and $Z$, with respective indices $a, b, c, d$ in Table~\ref{tele_index} and respective phase factors $E_Q$ and $E_P$ in
  Table~\ref{tele_phase}. Note that the $Y$ gate is defined in (\ref{Y_gate}). }
\end{center}
\end{table*}

\begin{table*}
\begin{center}
\footnotesize
\begin{tabular}{c|c|c|c}\hline\hline
  $B_{\textit{ell}}$ & \textit{CU} & $Q$ & $P$ \\ \hline
  & \textit{CNOT} & $E_QZ^aX^bZ^d$ & $E_PX^bX^cZ^d$\\ \cline{2-4}
  & \textit{CZ} & $E_QZ^aX^bZ^c$ & $E_PZ^bX^cZ^d$\\ \cline{2-4}
  & $C_H$ & $E_QX^aZ^bZ^d$ & $E_PX^aX^cZ^d$\\ \cline{2-4}
  & $C_H^{-1}$ & $E_QX^aZ^bX^d$ & $E_PX^bX^cZ^d$\\ \cline{2-4}
  & $B$ & $(-1)^{a}E_QY^aX^bX^cX^d$ & $(-1)^{d}E_PY^aZ^cX^d$ \\ \cline{2-4}
  \raisebox{1.6ex}[0pt]{$B/R$} & $B^{-1}$ & $E_QY^aX^bX^cX^d$ & $(-1)^cE_PY^aZ^cX^d$ \\ \cline{2-4}
  & $Q$ & $E_QX^aZ^bY^cY^d$ & $(-\sqrt{-1})^cE_PX^aX^bZ^cY^d$ \\ \cline{2-4}
  & $Q^{-1}$ & $(\sqrt{-1})^{a}(\sqrt{-1})^{b}E_QX^aZ^bY^d$ & $(\sqrt{-1})^{c}(\sqrt{-1})^{d}E_PY^aY^bY^cX^d$ \\ \cline{2-4}
  & $R$ & $E_QX^aX^bZ^cY^d$ & $ (\sqrt{-1})^cE_PX^aZ^bY^cY^d$ \\ \cline{2-4}
  & $R^{-1}$ & $(-\sqrt{-1})^{a}(-\sqrt{-1})^{b}E_QY^aZ^cX^d$ & $(-\sqrt{-1})^{c}(-\sqrt{-1})^{d}E_PX^aY^bY^cY^d$ \\ \hline\hline
\end{tabular}
\caption{\label{tele_qp_br} Local unitary operators $Q$ and $P$  in the teleportation equation
  (\ref{tele_two})  (or in Figure~\ref{two_tele}),  with
  the teleportation operator (\ref{tele_operator}) in terms of the Bell transforms $B$ and $R$,
  for two-qubit gates \textit{CU} including the \textit{CNOT} gate, the \textit{CZ} gate, the Bell transforms $B_{\textit{ell}}=C_H, B, Q, R$ and their inverses.
  Indices $a, b, c, d$ are shown in Table~\ref{tele_index}, and phase factors $E_Q$ and $E_P$ are in
  Table~\ref{tele_phase}. }
\end{center}
\end{table*}

\begin{table*}
\begin{center}
\footnotesize
\begin{tabular}{c|c|c|c|c}\hline\hline
   & $a$ & $b$ & $c$ & $d$ \\ \hline
  $C_H$ & $j_1+l_1$ & $i_1+k_1$ & $i_2+k_2$ & $j_2+l_2$ \\ \hline
  $B$ & $j_1+l_1$ & $i_1+j_1+k_1+l_1$ & $i_2+j_2+k_2+l_2$ & $j_2+l_2$ \\ \hline
  $Q$ & $i_1+j_1+k_1+l_1$ & $i_1+k_1$ & $i_2+k_2$ & $i_2+j_2+k_2+l_2$ \\ \hline
  $R$ & $i_1+k_1$ & $i_1+j_1+k_1+l_1$ & $i_2+j_2+k_2+l_2$ & $i_2+k_2$ \\ \hline\hline
\end{tabular}
\caption{\label{tele_index} The indices $a, b, c, d$ for the Bell transforms $B_{\textit{ell}}=C_H, Q$ in Table~\ref{tele_qp_cq}
and the Bell  transforms $B_{\textit{ell}}=B, R$ in Table~\ref{tele_qp_br}. All the index addition $+$ is the binary addition. }
\end{center}
\end{table*}


\begin{table*}
\begin{center}
\footnotesize
\begin{tabular}{c|c|c}\hline\hline
   & $E_Q$ & $E_P$ \\ \hline
  $C_H$ & $(-1)^{j_1\cdot k_1}$ & $(-1)^{i_2\cdot l_2}$ \\ \hline
  $B$ & $(-1)^{(k_1+l_1)\cdot (j_1+1)}$ & $(-1)^{(i_2+j_2)\cdot (l_2+1)}$ \\ \hline
  $Q$ & $(-1)^{(i_1+j_1)\cdot k_1}(-\sqrt{-1})^{j_1}(\sqrt{-1})^{l_1}$ & $(-1)^{(k_2+l_2)\cdot i_2}(-\sqrt{-1})^{j_2}(\sqrt{-1})^{l_2}$ \\ \hline
  $R$ & $(-1)^{(k_1+l_1)\cdot i_1}(-\sqrt{-1})^{l_1}(\sqrt{-1})^{j_1}$ & $(-1)^{(i_2+j_2)\cdot k_2}(-\sqrt{-1})^{l_2}(\sqrt{-1})^{j_2}$ \\ \hline\hline
  \end{tabular}
 \caption{\label{tele_phase}  The phase factors $E_Q$ and $E_P$ for the Bell transforms $B_{\textit{ell}}=C_H, Q$ in Table~\ref{tele_qp_cq}
and the Bell transforms $B_{\textit{ell}}=B, R$ in Table~\ref{tele_qp_br}. All the index multiplication $\cdot$
is the logical AND operation.  }
\end{center}
\end{table*}

We study the fault-tolerant construction of the universal quantum gate set using the teleportation operator (\ref{tele_operator}) or
(\ref{tele_operator2}). Refer to Subsection~\ref{universal gate set}, we know that an entangling two-qubit gate with all single-qubit
gates are capable of performing universal quantum computation. An entangling two-qubit Clifford
gate can be any one of the \textit{CNOT} gate, the \textit{CZ} gate, the Bell transforms $C_H$, $B$, $Q$, $R$ and
their inverses $C^{-1}_H$, $B^{-1}$, $Q^{-1}$, $R^{-1}$. Single-qubit gates can be generated by the Hadamard gate $H$ and the $T$ gate.
Note that all of these quantum gates have been defined in the previous sections and in the appendix.

To perform a single-qubit gate $U$ on the unknown qubit state $|\alpha\rangle$, namely $U|\alpha\rangle$, 
Alice prepares the quantum state given by
\eq
|\alpha\rangle \otimes |\psi_U (k^\prime,l^\prime)\rangle,
\en
with $|\psi_U (k^\prime,l^\prime)\rangle=(1\!\! 1_2 \otimes U)B_{\textit{ell}} |kl\rangle$
(different from $|\Psi_U\rangle$ (\ref{Psi_U})). Note that the bijective mappings between
$k^\prime, l^\prime$ and $k,l$ are defined in (\ref{index_function_p}). Then Alice applies the Bell measurement
denoted by $B^{-1}_{\textit{ell}}\otimes 1\!\! 1_2$ to the prepared
quantum state. These two successive operations lead to the teleportation equation given by
\eqa
\label{tele_single}
(B_{\textit{ell}}^{-1}\otimes 1\!\! 1_2) (1\!\! 1_2 \otimes 1\!\! 1_2 \otimes U ) (1\!\! 1_2 \otimes B_{\textit{ell}}) |\alpha\rangle|\textit{kl}\rangle
= \frac 1 2 \sum_{i,j=0}^{1} |i j \rangle S_{\textit{kl}} R_{\textit{ij}} U |\alpha\rangle,
\ena
with $S_{\textit{kl}} R_{\textit{ij}} = U (V_{\textit{kl}} U_{\textit{ij}} ) U^\dag$.  When Bob gets the classical two-bit message $(i, j)$
 from Alice, he performs the local unitary correction
operator $R_{\textit{ij}}^{-1} S_{\textit{kl}}^{-1}$ on his qubit to obtain the expected qubit state $U|\alpha\rangle$. Refer to
 Figure~\ref{single_tele} for the quantum circuit associated
with the teleportation equation (\ref{tele_single}).

Note that the $V_{\textit{kl}} U_{\textit{ij}}$ gate (Table \ref{tele_uv}) is a Pauli gate. As the single-qubit gate $U$ is the Hadamard gate $H$,
the $S_{\textit{kl}} R_{\textit{ij}}$ gate (Table \ref{tele_sr_h}) is still a Pauli gate. As the $U$ gate is the $T$ gate (\ref{T_gate}), the 
$S_{\textit{kl}} R_{\textit{ij}}$ gate (Table \ref{tele_sr_t}) is a
Clifford gate. Hence,  the fault-tolerant procedure of performing
the $T$ gate consists of two steps \cite{GC99}: The first step is to fault-tolerantly prepare the state $|\psi_U (k^\prime,l^\prime)\rangle$ with $U=T$, and
the second step is to fault-tolerantly perform the associated Clifford gate
$R_{\textit{ij}}^{-1} S_{\textit{kl}}^{-1}$.

The fault-tolerant construction of a two-qubit gate \textit{CU} depends on the fault-tolerant construction  of the quantum state given by
\eq
|\psi_{\textit{CU}}\rangle = (1\!\! 1_2 \otimes \textit{CU} \otimes 1\!\! 1_2  ) (B_{\textit{ell}} \otimes B_{\textit{ell}}) |k_1 l_1\rangle \otimes |k_2 l_2\rangle,
\en
which is a four-qubit state (different from $|\Psi_{\textit{CU}}\rangle$ (\ref{Psi_CU})). This state together with an unknown two-qubit product
state $|\alpha\beta\rangle$  given by
\eq
\label{state_prepared}
 |\alpha\rangle \otimes |\psi_{\textit{CU}}\rangle \otimes |\beta\rangle,
\en
is the prepared quantum state to be used. Applying the joint Bell measurement given by
\eq
 B^{-1}_{\textit{ell}} \otimes 1\!\! 1_2 \otimes 1\!\! 1_2 \otimes B^{-1}_{\textit{ell}},
\en
to the prepared quantum state (\ref{state_prepared}) gives rise to the teleportation equation
 \eqa
 \label{tele_two}
 & &  (B^{-1}_{\textit{ell}} \otimes 1\!\! 1_2 \otimes 1\!\! 1_2 \otimes B^{-1}_{\textit{ell}}) |\alpha\rangle \otimes |\psi_{\textit{CU}}\rangle \otimes |\beta\rangle
 \nonumber\\
 & & = \frac 1 4 \sum_{i_1, j_1=0}^1 \sum_{i_2, j_2=0}^1
 ( 1\!\! 1_2 \otimes 1\!\! 1_2 \otimes Q \otimes P \otimes 1\!\! 1_2 \otimes 1\!\! 1_2 )|i_1 j_1\rangle\otimes \textit{CU} |\alpha\beta\rangle\otimes|i_2 j_2\rangle,
 \ena
with $Q \otimes P$ defined by
\eq
\label{qp}
 Q \otimes P = \textit{CU} ( V_{\textit{k}_1 \textit{l}_1} U_{\textit{i}_1 \textit{j}_1} \otimes V^T_{\textit{k}_2 \textit{l}_2} U^T_{\textit{i}_2 \textit{j}_2}   ) 
  \textit{CU}^\dag,
\en
where the teleportation equations (\ref{tele_eq333}) and  (\ref{tele_eq444}) have been exploited. Hence the unitary correction operator is
given by $Q^\dag \otimes P^\dag$. Refer to Figure~\ref{two_tele} for the quantum circuit associated with the teleportation equation (\ref{tele_two})
and refer to Table~\ref{tele_qp_cq}, Table~\ref{tele_qp_br}, Table~\ref{tele_index} and Table~\ref{tele_phase} for the local unitary 
operators $Q$ and $P$ (\ref{qp}).

We draw Figures \ref{tele_circuit}--\ref{two_tele} and make  Tables \ref{tele_uv}--\ref{tele_phase} in order to present a detailed
account on the fault-tolerant construction of single-qubit gates and two-qubit gates in teleportation-based quantum computation \cite{GC99, Nielsen03, Leung04,ZZP15}
using  the Bell transform. These results show that teleportation-based computation using the Bell transform presents a  platform on
which quantum Clifford gate computation \cite{NC2011, Gottesman97}, quantum matchgate computation \cite{Valiant02,Knill01,JM08,RFSW10,BG11}, and
quantum computation using the Yang--Baxter gates \cite{KL04,ZJG06} can be performed.

\section{Concluding remarks}

\label{sec concluding}

Inspired by the quantum Fourier transform \cite{NC2011,Preskill97} and its application to quantum information and 
computation, we define the unitary basis transformation from the product basis to the GHZ basis \cite{GHZ89, GHZ90}
as the GHZ transform, with the Bell transform  as the simplest example. Since the GHZ states  are widely used in 
quantum information sicence, we expect that the GHZ transform plays the important roles in various topics of quantum 
information and computation. For example, in this paper, we clearly show that the teleportation operator using the 
Bell transform plays a crucial role in quantum teleportation \cite{BBCCJPW93, Vaidman94, BBC98, BDM00, Werner01,PEWFB15}
and teleportation-based quantum computation \cite{GC99, Nielsen03, Leung04, ZZP15}.

The following are remarks on possible further research topics. On the Bell transform, we study its generalized form as a function of parameters,
refer to the Yang--Baxter gate $B(x)$ \cite{KL04,ZJG06,Zhang07,YZG13} depending on the spectral parameter $x$. On quantum
teleportation, we apply the GHZ transform to multi-qubit teleportation \cite{CZG06,CZ09,ZLLF12} or quantum teleportation
via non-maximally entangling resources \cite{LLG00,AP02, GR06}. There remains a natural  question about the multi-qubit generalization
of the teleportation operators (\ref{tele_operator}) and (\ref{tele_operator2}).
On universal quantum computation, we study topological and algebraic aspects in the one-way quantum computation \cite{RB01,Jozsa05, CLN05}
using the GHZ transform. On quantum circuit models, we try to explore interesting quantum algorithms via the GHZ
transform, with the help of quantum algorithms \cite{DJ92,Simon94,Shor94,Kitaev95,Jozsa98} based on the quantum Fourier transform.

{\em Notes Added.} After this paper had been completed for some time, the authors have realized in their another paper \cite{ZZ16} that it is
natural and meaningful to generalize the definitions of  the GHZ transform (\ref{GHZ_transform}) (or the Bell transform (\ref{b_matrix_formulation})).
The generalized GHZ transform  $\widetilde{\textit{GHZ}}^{(n)}$ is defined as
\eqa
\label{generalized_GHZ transform}
\begin{split}
 \widetilde{\textit{GHZ}}^{(n)} &= \sum_{j_1,j_2,\ldots,j_n=0}^1\, e^{i\phi_{\textit{k}_1\textit{k}_2\ldots \textit{k}_n}}\,
 (S^{(1)}_{\textit{k}_1\textit{k}_2\ldots \textit{k}_n}\otimes S^{(2)}_{\textit{k}_1\textit{k}_2\ldots \textit{k}_n}\otimes
  \ldots \otimes S^{(n)}_{\textit{k}_1\textit{k}_2\ldots \textit{k}_n} ) \nonumber\\
   &  |G(k_1,k_2,\ldots,k_n)\rangle\langle j_1,j_2,\ldots,j_n|,
\end{split}
\ena
where  $k_{l}=k_{l}(j_1,j_2,\ldots,j_n)$, $l=1, \ldots, n$ are bijective functions of $j_1,j_2,\ldots,j_n$;
$e^{i \phi_{\textit{k}_1\textit{k}_2\ldots \textit{k}_n}}$ is the phase factor; and $S^{(l)}_{\textit{k}_1\textit{k}_2\ldots \textit{k}_n}$, $l=1, \ldots, n$
are single-qubit gates. Note that the generalized GHZ transform (\ref{generalized_GHZ transform}) differs from the GHZ transform (\ref{GHZ_transform}) because
the latter does not involve single-qubit gates $S^{(l)}_{\textit{k}_1\textit{k}_2\ldots \textit{k}_n}$. For example, the generalized Bell 
transform $\widetilde{B}_{\textit{ell}}$ is defined as
 \eq
 \widetilde{B}_{\textit{ell}}=\sum_{k^\prime, l^\prime=0}^1\,e^{i\phi_{\textit{kl}}}(S^{(1)}_{\textit{kl}}\otimes S^{(2)}_{\textit{kl}})|\psi(k,l)\rangle\langle
  k^\prime, l^\prime|,
\en
where $k=k(k^\prime,l^\prime)$ and $l=l(k^\prime,l^\prime)$ are bijective functions of $k^\prime$ and $l^\prime$, respectively;
$e^{i \phi_{\textit{kl}}}$ is the phase factor; and $S^{(1)}_{\textit{kl}}$ and $S^{(2)}_{\textit{kl}}$ are single-qubit gates.

\section*{Acknowledgement}
This work was supported by the starting Grant 273732 of Wuhan University, P. R. China and is supported by
the NSF of China (Grant No. 11574237 and 11547310).

\appendix

\section{The permutation gates and Clifford gates}

\label{appendix A}

In the defining relations of the GHZ transform (\ref{high_Bell}) and the Bell transform (\ref{b_transform}),
we introduce the permutation gates (\ref{P_n}) and (\ref{phase_permutation}), which are not much involved in the paper.
In this appendix, we perform a further study on the permutation gates. First, we verify the two-qubit permutation
gate (\ref{phase_permutation}) as a Clifford gate. Second, we explain with examples that a multi-qubit permutation
gate (\ref{P_n}) is usually not a Clifford gate.

The permutation group $S_{2^n}$  is the set of all permutations of $2^n$ elements \cite{NC2011}, and it is
generated by transpositions $(J,J+1)$ with $1\le J \le 2^n-1$.  As usual, we study algebraic properties
of the transposition gates $T^{(n)}_{J,J+1}$ to understand the permutation gate (\ref{P_n}). Note that
the permutation gate (\ref{P_n})  forms a unitary representation of the permutation group $S_{2^n}$.
When the $n$-qubit product basis $|j_1 j_2\ldots j_n\rangle$ is relabeled as $|J\rangle$ given by
\eq
J= 2^{n-1} \cdot j_1 + 2^{n-2}\cdot j_2 + \ldots + j_n +1,
\en
with decimal addition, a unitary representation $T^{(n)}_{J,J+1}$ associated
with the transposition $(J,J+1)$ is given by $T^{(n)}_{J,J+1}|J\rangle=|J+1\rangle$.

The two-qubit permutation gate  (\ref{phase_permutation})  forms a representation of the permutation
group $S_4$  of four elements, and the associated transposition gates are respectively
denoted by  $T^{(2)}_{12}$, $T^{(2)}_{23}$ and $T^{(2)}_{34}$. The transposition gate $T^{(2)}_{12}$ defined
by $T^{(2)}_{12}|1\rangle=|2\rangle$ has the form
\eq
 T^{(2)}_{12} |ij\rangle =|i,i+j+1\rangle,
\en
so that the $T^{(2)}_{12}$ gate has the form
\eq
T^{(2)}_{12}=X_2 \textit{CNOT}_{12},
\en
with $X=H S^2 H$. The transposition gate $T^{(2)}_{23}$ defined by $T^{(2)}_{23} |2\rangle =|3\rangle$ has the
form of the \textit{SWAP} gate \cite{NC2011} as a product of three \textit{CNOT} gates,
\eq
\label{swap}
T^{(2)}_{23}=\textit{SWAP}=\textit{CNOT}_{12} \textit{CNOT}_{21} \textit{CNOT}_{12}.
\en
The transposition gate $T^{(2)}_{34}$ given by $T^{(2)}_{34}|3\rangle=|4\rangle$ has the form $T^{(2)}_{34}=\textit{CNOT}_{12}$.
Since three transposition gates $T^{(2)}_{12}$, $T^{(2)}_{23}$ and $T^{(2)}_{34}$ are Clifford gates \cite{NC2011, Gottesman97},
the permutation gate (\ref{phase_permutation}) generated by them is certainly the Clifford gate.

\begin{table*}
\begin{center}
\footnotesize
\begin{tabular}{c|c|c}
\hline\hline
Operation & Input & Output \\ \hline
 & $X_1$ & $X_1 \textit{CNOT}_{23}$ \\
 & $X_2$ & $X_2 \textit{CNOT}_{13}$ \\
 & $X_3$ & $X_3$ \\ \cline{2-3}
\raisebox{1.6ex}[0pt]{Toffoli gate} & $Z_1$ & $Z_1$ \\
 & $Z_2$ & $Z_2$ \\
 & $Z_3$ & $Z_3\textit{CZ}_{12}$ \\ \hline
 & $X_1$ & $X_1SWAP_{23}$ \\
 & $X_2$ & $X_2\textit{CNOT}_{13}\textit{CNOT}_{12}$ \\
 & $X_3$ & $X_3\textit{CNOT}_{13}\textit{CNOT}_{12}$ \\ \cline{2-3}
\raisebox{1.6ex}[0pt]{Fredkin gate} & $Z_1$ & $Z_1$ \\
 & $Z_2$ & $Z_2\textit{CZ}_{13}\textit{CZ}_{12}$ \\
 & $Z_3$ & $Z_3\textit{CZ}_{13}\textit{CZ}_{12}$ \\\hline\hline
\end{tabular}
\end{center}
\caption{\label{T F} Transformation properties of elements of the Pauli group ${\mathcal P}_3$ under conjugation by the Toffoli gate and Fredkin
gate, respectively. For example, $(\textit{Toffoli})\, Z_1\, (\textit{Toffoli})^\dag = Z_1$. The \textit{SWAP} gate is defined in (\ref{swap}).  }
\end{table*}

A three-qubit permutation gate (\ref{P_n}) may not be a Clifford gate because two three-qubit transposition gates $T^{(3)}_{67}$ and $T^{(3)}_{78}$
are not Clifford gates.  The  transposition gate $T^{(3)}_{67}$ is the Fredkin gate \cite{NC2011} given by
\eq
T^{(3)}_{67}|j_1j_2j_3\rangle=|j_1,j_1\cdot(j_2+j_3)+j_2,j_1\cdot(j_2+j_3)+j_3\rangle,
\en
which denotes the permutation between the product states $|101\rangle$ and $|110\rangle$.
The other  transposition gate $T^{(3)}_{78}$ is the Toffoli gate \cite{NC2011} given by
\eq
T^{(3)}_{78}|j_1j_2j_3\rangle=|j_1,j_2,j_1\cdot j_2+j_3\rangle,
\en
which denotes the permutation between the product states $|110\rangle$ and $|111\rangle$. It is well-known that
the Fredkin gate and the Toffoli gate are not Clifford gates, refer to Table~\ref{T F} for transformation properties of the elements of the Pauli group
${\mathcal P}_3$ under conjugation by the Toffoli gate and Fredkin gate, respectively.

In accordance with the definition of the controlled operation \cite{NC2011}, the Fredkin gate and Toffoli gate can be respectively viewed as the controlled
\textit{SWAP} gate and the controlled \textit{CNOT} gate \cite{NC2011}. Hence we introduce the controlled controlled \textit{SWAP} gate to denote
a four-qubit transposition gate $T^{(4)}_{14,15}$ and the controlled controlled \textit{CNOT} gate to denote a four-qubit transposition gate $T^{(4)}_{15,16}$. 
Both four-qubit transposition gates, $T^{(4)}_{14,15}$ and $T^{(4)}_{15,16}$,  are not Clifford gates, so a four-qubit permutation gate (\ref{P_n}) is not a 
Clifford gate. Similarly, with a series of controlled operations on the
\textit{SWAP} gate and the \textit{CNOT} gate, we can respectively construct the $n$-qubit transposition gates $T^{(n)}_{2^n-2,2^n-1}$ 
and $T^{(n)}_{2^n-1,2^n}$, which are not Clifford gates, so that an $n$-qubit ($n\ge3$) permutation gate $P^{(n)}$ (\ref{P_n}) may not 
be a Clifford gate in general.

\section{Notes on representative examples for the Bell transform}

\label{appendix B}

 In Section \ref{repr_examples}, we study the definition of the Bell transform with representative examples including the $C_H$ gate (\ref{ch_matrix}),
 the Yang--Baxter gate $B$ (\ref{b_matrix}), and the magic gates $Q$ (\ref{q_matrix}) and $R$ (\ref{r_matrix}). Here we verify these representative gates 
 and their inverses as maximally entangling Clifford gates and study exponential formulations of the $B$, $Q$, and $R$ gates with associated two-qubit 
 Hamiltonians.

\begin{table}
\begin{center}
\footnotesize
\begin{tabular}{c|c|c}
\hline\hline
Operation & Input & Output \\ \hline
  & $X_1$ & $Z_1$ \\
  & $X_2$ & $X_2$ \\
  \raisebox{1.8ex}[0pt]{$C_H$} & $Z_1$ & $X_1X_2$ \\
  & $Z_2$ & $Z_1Z_2$ \\  \hline
  & $X_1$ & $X_1$ \\
  & $X_2$ & $X_1Z_2$ \\
  \raisebox{1.8ex}[0pt]{$B$} & $Z_1$ & $-Y_1Y_2$ \\
  & $Z_2$ & $-X_1X_2$ \\  \hline
  & $X_1$ & $Z_1X_2$ \\
  & $X_2$ & $-i Y_1Z_2$ \\
  \raisebox{1.8ex}[0pt]{$Q$} & $Z_1$ & $X_1X_2$ \\
  & $Z_2$ & $Y_1Y_2$ \\  \hline
  & $X_1$ & $X_1Z_2$ \\
  & $X_2$ & $i Z_1Y_2$ \\
\raisebox{1.8ex}[0pt]{$R$} & $Z_1$ & $X_1X_2$ \\
 & $Z_2$ & $Y_1Y_2$ \\  \hline\hline
\end{tabular}
\caption{\label{clifford_bell}  Transformation properties of elements of the Pauli group ${\mathcal P}_2$ under conjugation
 by the Bell transforms $B_{\textit{ell}}=C_H, B, Q, R$. For example, $C_H X_1 C_H^\dag=Z_1$.
    The symbol $i$ denotes the imaginary unit, and the $Y$ gate is defined in (\ref{Y_gate}).  }
\end{center}
\end{table}

We recognize the Bell transforms $C_H$, $B$, $Q$ and $R$  as
Clifford  gates \cite{NC2011, Gottesman97}. For example, the elements of the Pauli group ${\mathcal P}_2$ on two qubits are transformed under
conjugation by the Yang--Baxter gate $B$ in the way
\eqa
\begin{aligned} B X_1 B^\dag &= X_1, & B X_2 B^\dag &=X_1 Z_2, \\ B Z_1 B^\dag &= -Y_1 Y_2, &  B Z_2 B^\dag &=- X_1 X_2, \end{aligned}
\ena
with the $Y$ gate defined in (\ref{Y_gate}), and thus the Yang--Baxter gate $B$ is a Clifford gate preserving the Pauli group 
under conjugation. Refer to Table~\ref{clifford_bell} for transformation properties of $X_1$, $X_2$, $Z_1$ and $Z_2$ under conjugation 
by the Bell transforms $B_{\textit{ell}}=C_H, B, Q, R$. Therefore, the Bell transforms $B, Q, R$ can be respectively formulated as products 
of the \textit{CNOT} gate, the $H$ gate and the phase gate $S$. The results are given by
 \eqa
 \begin{aligned}
  B &= \textit{CNOT}_{12} \, H_1X_1\textit{CNOT}_{21}\textit{CZ}_{12}\textit{CNOT}_{21}, \\
  Q &= \textit{CNOT}_{12}\, H_1\textit{CNOT}_{12}S_2,\\
  R &= \textit{CNOT}_{12}\, H_1S_1S_2\textit{CNOT}_{12},
 \end{aligned}
  \ena
where $\textit{CNOT}_{12} \, H_1=(C_H)_{12}$ and  the \textit{CZ} gate has the form of
\eq
 \textit{CZ}_{21}=(H \otimes 1\!\! 1_2) \textit{CNOT}_{21} (H\otimes 1\!\! 1_2),
\en
with $\textit{CZ}_{12}=\textit{CZ}_{21}$. Furthermore, with the research work \cite{RFSW10} by Ramelow et al., the
parity-preserving gate $G=G(A_G, B_G)$ (\ref{matchgate}) is reformulated as
\eq
G(A_G, B_G) = \textit{CNOT}_{12}\, \textit{CU}_{21}\, (A_G \otimes 1\!\! 1_2) \textit{CNOT}_{12},
\en
where the controlled-$U$ gate $\textit{CU}_{21}$ given by
\eq
 \textit{CU}_{21}=1\!\! 1_2\otimes |0\rangle\langle 0| + B_G A_G^{-1} \otimes |1\rangle\langle 1|,
\en
can be further decomposed as a tensor product of \textit{CNOT} gates and single-qubit gates, refer to Nielsen and Chuang's description on
controlled operations \cite{NC2011}. The Yang--Baxter gate $B$ has the form
\eq
B = \textit{CNOT}_{12} \textit{CZ}_{21} (Z\,H \otimes 1\!\! 1_2) \textit{CZ}_{21}  \textit{CNOT}_{12},
\en
or equivalently
\eq
 B = \textit{CNOT}_{12} (Z\, H\otimes 1\!\! 1_2) \textit{CNOT}_{21} \textit{CZ}_{21}  \textit{CNOT}_{12},
\en
which has a more simplified form
\eq
\label{clifford_ybe}
B = \textit{CNOT}_{21} (1\!\! 1_2 \otimes Z\,H) \textit{CNOT}_{21}.
\en
The Bell transforms $Q$ and $R$ have the decomposition such as (\ref{clifford_ybe}), respectively, given by
 \eqa
 \begin{aligned}
 Q &= \textit{CNOT}_{12}( H\,S \otimes  S) \textit{CZ}_{21} \textit{CNOT}_{12}, \\
 R &= \textit{CNOT}_{12} (H\,S^\dag\otimes S^\dag ) \textit{CNOT}_{12},
 \end{aligned}
 \ena
with $S^\dag =S Z$.

The Bell transforms $C_H$, $B$, $Q$ and $R$ are Clifford gates, so their inverses $C^{-1}_H$, $B^{-1}$, $Q^{-1}$ and $R^{-1}$ are also
Clifford gates, refer to Table~\ref{clifford_bell_in}, for transformation properties of generators of the Pauli group  ${\mathcal P}_2$
under conjugation by $C^{-1}_H$, $B^{-1}$, $Q^{-1}$ and $R^{-1}$, respectively. According to the description of quantum teleportation
in Section~\ref{sec: teleportation} that the Bell transform (\ref{def_bell}) is explained as the creation operator
of Bell states and the inverse of the Bell transform is associated with Bell measurements,
the inverse of the Bell transform is not the Bell transform in general. We derive the explicit forms of  $C^{-1}_H$, $B^{-1}$, $Q^{-1}$
and $R^{-1}$ in the following. The inverse of the Bell transform $C_H$ (\ref{ch_matrix})  has the form
\eq
C_H^{-1}=\frac{1}{\sqrt{2}}\left(
               \begin{array}{cccc}
                 1 & 0 & 0 & 1 \\
                 0 & 1 & 1 & 0 \\
                 1 & 0 & 0 & -1 \\
                 0 & 1 & -1 & 0 \\
               \end{array}
             \right),
\en
which gives rise to $C_H^{-1}|00\rangle ={\frac 1 {\sqrt 2}} (|00\rangle+|01\rangle)$, so the $C_H^{-1}$ gate is not the Bell transform.
The inverse of the magic gate $Q$ (\ref{q_matrix}) has the form
\eq
  Q^{-1}=\frac{1}{\sqrt{2}}\left(
                      \begin{array}{cccc}
                        1 & 0 & 0 & 1 \\
                        0 & -i & -i & 0 \\
                        0 & 1 & -1 & 0 \\
                        -i & 0 & 0 & i \\
                      \end{array}
                    \right),
\en
which leads to $Q^{-1}|11\rangle=(1\!\! 1_2\otimes S)|\psi(1,1)\rangle$, and the inverse of the magic gate $R$ (\ref{r_matrix}) given
by
\eq
 R^{-1}=\frac{1}{\sqrt{2}}\left(
    \begin{array}{cccc}
     1 & 0 & 0 & 1 \\
       0 & i & i & 0 \\
         0 & -1 & 1 & 0 \\
          i & 0 & 0 & -i \\
          \end{array}
     \right),
\en
has $R^{-1}|00\rangle=Q^{-1}|11\rangle$. So the $Q^{-1}$ and $R^{-1}$ gates are not the Bell transform. Note that the $Q^{-1}$ gate is
a matchgate and the $R^{-1}$ gate is a parity-preserving non-matchgate. Occasionally, the inverse of the Yang--Baxter gate $B$ (\ref{b_matrix})
given by
\eq
B^{-1}=\frac{1}{\sqrt{2}}\left(
                      \begin{array}{cccc}
                        1 & 0 & 0 & -1 \\
                        0 & 1 & 1 & 0 \\
                        0 & -1 & 1 & 0 \\
                        1 & 0 & 0 & 1 \\
                      \end{array}
                    \right),
\en
is the Bell transform and the Yang--Baxter gate \cite{KL04,ZJG06}.

\begin{table}
\begin{center}
\footnotesize
\begin{tabular}{c|c|c}
\hline\hline
Operation & Input & Output \\ \hline
  & $X_1$ & $Z_1X_2$ \\
  & $X_2$ & $X_2$ \\
  \raisebox{1.8ex}[0pt]{$C_H^{-1}$} & $Z_1$ & $X_1$ \\
  & $Z_2$ & $X_1Z_2$ \\  \hline
  & $X_1$ & $X_1$ \\
  & $X_2$ & $-X_1Z_2$ \\
  \raisebox{1.8ex}[0pt]{$B^{-1}$} & $Z_1$ & $Y_1Y_2$ \\
  & $Z_2$ & $X_1X_2$ \\  \hline
  & $X_1$ & $i Z_1Y_2$ \\
  & $X_2$ & $i Y_2$ \\
  \raisebox{1.8ex}[0pt]{$Q^{-1}$} & $Z_1$ & $i X_1Y_2$ \\
  & $Z_2$ & $i Y_1X_2$ \\  \hline
  & $X_1$ & $-i Y_2$ \\
  & $X_2$ & $-i Z_1Y_2$ \\
\raisebox{1.8ex}[0pt]{$R^{-1}$} & $Z_1$ & $-i Y_1X_2$ \\
 & $Z_2$ & $-i X_1Y_2$ \\  \hline\hline
\end{tabular}
\caption{\label{clifford_bell_in} Transformation properties of elements of the Pauli group ${\mathcal P}_2$ under conjugation
 by the inverses of the Bell transforms $B^{-1}_{\textit{ell}}=C_H^{-1}, B^{-1}, Q^{-1}, R^{-1}$.  For example, $C_H^{-1} X_2 C_H =X_2$.}
\end{center}
\end{table}

The Bell transform and its inverse are maximally entangling two-qubit gates because the product states are separable states and the
Bell states are maximally entangled states in any entanglement measurement theory \cite{PV05, Bruss02}. We calculate
the entangling powers \cite{ZZF00} of the Bell transforms $C_H, B, Q, R$ and their inverses to support this statement. Any two-qubit gate
$U$ \cite{KC02} is locally equivalent to a two-qubit gate
$e^{i (a X\otimes X + b Y\otimes Y + c Z\otimes Z)}$ with three non-local parameters $(a, b, c)$, and the entangling power $e_p(U)$ \cite{BG11}
of this two-qubit gate $U$ has the form
\eq
\label{entangling power}
e_p(U)=1- \cos^2 2a \cos^2 2b \cos^2 2c - \sin^2 2a \sin^2 2b \sin^2 2c,
\en
with the maximum 1. The non-local parameters $(a, b, c)$ of the Bell transform $C_H$ and its inverse $C_H^{-1}$ are the same as
those of the \textit{CNOT} gate, which is $(\frac \pi 4, 0, 0)$. After some algebra, those of the Yang--Baxter
gate $B$ and its inverse $B^{-1}$ are $(\frac \pi 4, 0, 0)$.
The magic gate $Q$ and its inverse $Q^{-1}$ are locally equivalent to the inverse of the Yang--Baxter gate
$B^\prime=e^{\frac \pi 4 Y \otimes X}$ (\ref{B'})
with $Q=(B^\prime)^{-1} (Z\otimes S)$ or $Q^{-1}=(1\!\! 1_2 \otimes S^\dag ) (B^\prime)^{-1} (Z\otimes 1\!\! 1_2)$, so
that the gates $Q$, $Q^{-1}$ and $(B^\prime)^{-1}$ have the same non-local parameters. Note that the $(B^\prime)^{-1}$ gate has non-local
parameters $(\frac \pi 4, 0, 0)$. The magic gate $R$  and its inverse $R^{-1}$ are also associated with
 $(B^\prime)^{-1}$ in the way
\eqa
\begin{aligned}
 R &= e^{-i \frac \pi 4} e^{i \frac \pi 4 Z\otimes Z} (B^\prime)^{-1} (S\otimes 1\!\! 1_2 ), \\
 R^{-1} &= (S\otimes 1\!\! 1_2) (B^\prime)^{-1}  e^{i \frac \pi 4 Z\otimes Z}  (1\!\! 1_2 \otimes Z ),
 \end{aligned}
\ena
which give non-local parameters of the $R$ and $R^{-1}$ gates as  $(\frac \pi 4, 0, \frac \pi 4)$. With the formula (\ref{entangling power}),
the entangling power of the Bell transforms $C_H, B, Q, R$ and their inverses can be calculated exactly as 1, so all of them
are maximally entangling gates.

 We study how to prepare the Bell transforms $C_H$, $B$,  $Q$ and $R$ and their inverses in experiments.  They are Clifford gates so
 they can be generated by the elementary Clifford gates which are the ordinary quantum gates in experiments \cite{NC2011}; for example,
 the $C_H$ gate is easily performed as a tensor product of the \textit{CNOT} gate and the Hadamard gate $H$. On the other hand,
 we study the exponential formulations of three parity-preserving
 gates $B$, $Q$ and $R$ with associated two-qubit Hamiltonians, and with the results we discuss the essential difference
 between the matchgates $B, Q$ and the non-matchgate $R$ from the viewpoint of universal quantum computation.
 Given the Hamiltonian $H_B=i X\otimes  Y$,  the Yang--Baxter gate $B$ has the form
  \eq
  B=e^{-iH_B\,\, t}|_{t=\frac \pi 4},
  \en
where $t$ denotes the evolutional time. The magic gate $Q$  has the exponential form  with the global phase $e^{i 3 \pi/4}$ given by
  \eq
  Q=e^{i\frac{3\pi}{4}}e^{-\frac{\pi}{4} Y\otimes X}e^{-i\frac{\pi}{4}(2Z\otimes 1\!\!1_2+1\!\!1_2\otimes Z)},
  \en
 which gives rise to a time-dependent Hamiltonian,
  \eq
  H_Q(t) =  \theta(\frac{\pi}{4}-t) (2Z\otimes {1\!\!1_2}+{1\!\!1_2} \otimes Z) + \theta(t-\frac{\pi}{4})  (-i Y\otimes X),
  \en
  with the step  functions $\theta(\frac{\pi}{4}-t)$ and $\theta(t-\frac{\pi}{4})$ and $0\leq t \leq \pi/2$.
  Equivalently, the magic gate $Q$ has the other exponential formulation
  \eq
  Q=e^{-i\frac{\pi}{4}}e^{-i\frac{\pi}{4}(-2X\otimes X + Y\otimes  Y)}e^{-\frac{\pi}{4}( Y\otimes X)},
  \en
  where the associated Hamiltonian has the form
  \eq
  H^\prime_Q(t)=\theta(\frac{\pi}{4}-t)(-i Y\otimes X) + \theta(t-\frac{\pi}{4})(-2X\otimes X + Y\otimes  Y),
  \en
 with $t\in [0, \pi/2]$. The magic gate $R$  has the exponential form given by
  \eq
  R=e^{-i\frac{\pi}{4}(-i Y\otimes X-Z\otimes Z)}e^{-i\frac{\pi}{4}(Z\otimes 1\!\!1_2)},
  \en
with a time-dependent Hamiltonian given by
  \eq
  H_R(t)= \theta(\frac{\pi}{4}-t) (Z\otimes 1\!\!1_2) + \theta(t-\frac{\pi}{4}) (-i Y\otimes X-Z\otimes Z),
  \en
and  has the other equivalent exponential form
  \eq
  R=e^{-i\frac{\pi}{4}(X\otimes X-Z\otimes Z)}e^{-\frac{\pi}{4}( Y\otimes X)},
  \en
  with the associated Hamiltonian
  \eq
  H^\prime_R(t)= \theta(\frac{\pi}{4}-t) (-i Y\otimes X)  + \theta(t-\frac{\pi}{4}) (X\otimes X-Z\otimes Z),
  \en
 with $t\in [0, \pi/2]$. Similarly, we can derive the exponential formulations of the inverses of the Bell transforms $B^{-1}, Q^{-1}$ and $R^{-1}$
  with associated two-qubit Hamiltonians, respectively, for example, $B^{-1}=-e^{-i H_B t}|_{t=3 \pi /4}$.
  Note that a two-qubit matchgate \cite{Knill01} is generated by a Hamiltonian as a linear combination of $Z\otimes 1\!\! 1_2$, $1\!\! 1_2 \otimes Z$,
  $X\otimes X$, $Y\otimes Y$, $X\otimes Y$ and $Y\otimes X$. Among the above Hamiltonians, only the Hamiltonians of the magic gate
  $R$ has an exceptional term $Z\otimes Z$, so the $B$ gate and $Q$ gate are matchgates and the $R$ gate is a non-matchgate.  Quantum computation with
  matchgates $B$ or $Q$ can be efficiently simulated on a classical computer, whereas quantum computation with the parity-preserving gate $R$ can boost
  universal quantum computation mainly due to the computational power of the term $\exp^{i \pi/4 (Z\otimes Z)}$, refer to \cite{BG11, BK02}.

\end{document}